\documentclass{iopart}

\usepackage{iopams}
\usepackage{setstack}
\usepackage{graphicx}
\usepackage{diagrams}
\usepackage{pst-node,pst-tree}
\usepackage{theorem}

\theorembodyfont{\upshape}\theoremheaderfont{\sffamily\bfseries}
\newtheorem{theo}{Theorem}
\newtheorem{prop}{Proposition}
\newtheorem{rec}{Recursion}
\newtheorem{defi}{Definition}
\newtheorem{coro}{Corollary}

\newcommand{\bbdot}{\mbox{$^{\boldsymbol{.}}$}}

\begin{document}

\jl{1}

\title[Cluster density functional theory for lattice models]%
{Cluster density functional theory for lattice models
based on the theory of M\"obius functions}

\author{Luis Lafuente and Jos\'e A Cuesta}

\address{Grupo Interdisciplinar de Sistemas Complejos (GISC),
Departamento de Matem\'aticas, Universidad Carlos III de Madrid,
28911 Legan\'es, Madrid, Spain}

\eads{\mailto{llafuent@math.uc3m.es} and \mailto{cuesta@math.uc3m.es}}

\begin{abstract}
Rosenfeld's fundamental measure theory for lattice models is
given a rigorous formulation in terms of the theory of M\"obius
functions of partially ordered sets. The free-energy density
functional is expressed as an expansion in a finite set of
lattice clusters. This set is endowed a partial order, so that the
coefficients of the cluster expansion are connected to its
M\"obius function. Because of this, it is rigorously proven that
a unique such expansion exists for any lattice model. The
low-density analysis of the free-energy functional motivates a
redefinition of the basic clusters (zero-dimensional cavities)
which guarantees a correct zero-density limit of the pair and
triplet direct correlation functions.
This new definition extends Rosenfeld's theory to lattice
model with any kind of short-range interaction (repulsive
or attractive, hard or soft, one- or multi-component\dots).
Finally, a proof is given that these functionals have a consistent
dimensional reduction, i.e.\ the functional for dimension
$d'$ can be obtained from that for dimension $d$ ($d'<d$) if the
latter is evaluated at a density profile confined to a
$d'$-dimensional subset.
\end{abstract}

\pacs{61.20.Gy, 05.20.Jj, 05.50.+q, 02.10.Ox}
\ams{82B05, 82B20, 06A07}

\submitted


\eqnobysec

\section{Introduction}
\label{sec1}
Rosenfeld's fundamental measure theory (FMT) is a singularity in the
world of approximate density functional theories. While all approximate
functionals are built aiming at incorporating as much information
on the uniform phase as there is available (Evans~1992), FMT is
constructed on purely geometrical arguments (Rosenfeld~1989). Because
of this, typical density functional recipes provide simple functionals
with great flexibility to incorporate data of very different nature
on the thermodynamics and the structure of the fluids, while 
fundamental-measure (FM) functionals have a very rigid structure 
which rejects almost any deviation from orthodoxy (Tarazona~2002,
Cuesta \mbox{\etal{}2002)}. The counterpart of this is that FM functionals 
exhibit a set of very special properties not shared by any other
approximate functional. To begin with, the structure of the fluids
is \textsl{predicted} rather than input (as in the other theories).
Furthermore, FMT is naturally formulated for multicomponent systems,
while other theories have serious difficulties to pass from one-component
fluids to even binary mixtures (Denton and Ashcroft~1991, Choudhury
\mbox{\etal{}2002)}. But perhaps the most striking and characteristic property
which distinguishes FM functionals is \textsl{dimensional reduction}.
This means that if a $d$-dimensional system is constrained to lie
in a $d'$-dimensional subset (with $d'<d$) and we evaluate the
$d$-dimensional FM functional at the density profile describing
this $d'$-dimensional confinement, then we obtain as a result the
FM functional for the $d'$-dimensional system (Rosenfeld \mbox{\etal{}1996,}
Rosenfeld \mbox{\etal{}1997,} Cuesta and Mart\'{\i}nez-Rat\'on~1997a,
Cuesta and Mart\'{\i}nez-Rat\'on~1997b). This extraordinary
consistency of the theory (as a matter of fact at the origin
of its rigidity (Cuesta \mbox{\etal{}2002))} cannot be found in any other
density functional theory, and is obviously a feature that exact
functionals possess. Dimensional reduction can be extended down
to zero-dimensional (0D) confinements (cavities holding no more
than a particle), and as theory has developed, it has become
more and more clear that this latter fact can be reformulated as
a constructing principle of any FM functional (Tarazona and
Rosenfeld~1997, Cuesta and Mart\'{\i}nez-Rat\'on~1997a,
Tarazona~2000).

FMT was first formulated as a continuum theory, but in a series
of recent works the authors have extended the theory to lattice
models and formulated a lattice fundamental measure theory (LFMT)
(Lafuente and Cuesta~2002, Lafuente and Cuesta~2003,
Lafuente and Cuesta~2004, Lafuente~2004).
The extension has been most revealing about the structure of
FM functionals. In fact, LFMT, whose roots are very close to
the continuum theory (Lafuente and Cuesta~2002), has its most
natural formulation as a cluster theory, becoming a kind of
density functional version of Kikuchi's cluster variation method
(Kikuchi~1951) in Morita's formulation (Morita~1994). In its
latest formulation, LFMT can be constructed, for hard-particle
models, out of the exact free-energy functional of a given set of
0D cavities (Lafuente and Cuesta~2004, Lafuente~2004).
The final result is of
an extraordinary simplicity, given the fact that it provides
the exact functional of many one-dimensional systems, and a good
approximation for higher-dimensional ones (typically a Bethe
approximation for lattice gases with nearest-neighbor exclusion).
On the other hand, LFMT exhibits dimensional-reduction consistency
even beyond dimensional reduction (that is, we remain within
LFMT under transformations more general than lower-dimensional
confinements of the system; see (Lafuente and Cuesta~2003) for
some examples, as well as \sref{sec:sec5a} of this article).

In this work we will present a formulation of LFMT based
on a powerful combinatorial tool known as the \textsl{theory of
M\"obius functions} (Rota~1964, Aigner~1979, Stanley~1999), whose
natural context are \textsl{incidence algebras} defined on partially
ordered sets (or \textsl{posets).} This formalism will allow
us to rigorously prove a list of results about LFMT. Thus,
after revisiting LFTM as reported in (Lafuente and Cuesta~2004)
and analyzing its structure (\sref{sec2}), we will prove
(\sref{sec:sec3}) that given a lattice model as well as a
basic set of clusters $\mathcal{W}_{\mathrm{max}}$ (to be
precisely defined later) of the lattice, there exists a unique
linear combination of the free-energy density functionals
on subcluster of $\mathcal{W}_{\mathrm{max}}$ which yields the
exact free energy when evaluated at 0D density profiles
(i.e.\ density profiles with support one of those clusters).
A special choice of $\mathcal{W}_{\mathrm{max}}$ gives rise to
LFMT (\sref{sec:sec4}), but the cluster expansion has a wider
range of applications (\sref{sec:sec4b}). We study the low-density
limit of the cluster expansion and redefine the clusters of
LFMT (0D cavities) in such a way that for any model it is guaranteed
that the zero-density limits of the pair and triplet direct
correlation functions are exact (\sref{sec:sec5}), and suggest
how to extend the definition in order to systematically incorporate
higher order direct correlation functions. Finally, the behaviour
of the cluster expansion under the action of a mapping is analyzed,
a consequence of which is the proof that LFMT is closed under
dimensional reduction and other more general mappings
(\sref{sec:sec5a}).

The basic result used in the above proofs is a theorem about
the M\"obius function which is stated and proven in an appendix.
The appendix also contains an important result of the theory
of M\"obius functions (the cross-cut theorem) that yields some
simplifications in the calculation of the M\"obius functions of
a given poset. We conclude this article with a discussion about
some of the consequences of this reformulation of LFMT (the
simplicity of its application being perhaps one of the most
remarkable because of its practical consequences) as well as
some open questions related to this theory.

\section{Lattice fundamental measure recipe reviewed: a
multicomponent example}
\label{sec2}
In this section we will review the procedure recently proposed
by the authors (Lafuente and Cuesta~2004) to construct a FM
functional for any hard-core lattice model. As in the latest
versions for continuum models (Tarazona and Rosenfeld 1997,
Tarazona 2000), the constructive principle
is based on the exact dimensional crossover to 0D cavities.
In brief, the aim of this procedure is to build the
simplest functional (under certain assumptions) which
applied to 0D cavities, produces the exact result.

The first hypothesis of the recipe, based on the exact
functional for one-dimensional hard rods (Lafuente and
Cuesta~2002) and the common pattern shared by all lattice FM
functionals studied by the authors
(Lafuente and Cuesta~2002, 2003), is that the excess (over ideal)
free-energy functional of an arbitrary hard-core multicomponent system
in a lattice $\mathcal{L}$ has the form
(in units of $kT$, the Boltzmann constant times the temperature)
\begin{equation}
\label{ec:generalfunctional}
\fl
\mathcal{F}^{\mathrm{ex}}_\mathrm{FM}[\brho]=\sum_{\bi{s}\in\mathcal{L}}
\sum_{k\in\mathcal{I}} a_k \Phi_0\bigl(n^{(k)}(\bi{s})\bigr),\qquad
n^{(k)}(\bi{s})\equiv\sum_{i=1}^p\sum_{\bi{t}\in
\mathcal{C}^{(k)}_i(\bi{s})}\rho_i(\bi{t}),
\end{equation}
where $\mathcal{I}$ is a set of indices which label the different
weighted densities $n^{(k)}(\bi{s})$; $a_k$ are integer coefficients
which depend on the specific model; $\Phi_0(\eta)\equiv\eta+(1-\eta)
\ln(1-\eta)$ is the excess free energy of a cavity admitting no
more than one particle in which $0\leq \eta \leq 1$ is the
average occupancy; $\brho(\bi{s})=\bigl(\rho_1(\bi{s}),\ldots,
\rho_p(\bi{s})\bigr)$ denotes the vector of one-particle density
functions of the different $p$ species, and
$\boldsymbol{\mathcal{C}}^{(k)}(\bi{s})
\equiv\bigl(\mathcal{C}^{(k)}_1(\bi{s}),\ldots,
\mathcal{C}^{(k)}_p(\bi{s})\bigr)$ are vectors formed by subsets of
lattice nodes,
i.e.\ $\mathcal{C}^{(k)}_i(\bi{s})\subset\mathcal{L}$
($i$ is a species subindex). Hereafter,
$\boldsymbol{\mathcal{C}}^{(k)}(\bi{s})$ will be referred to as
(multicomponent) \textsl{cluster} or \textsl{cavity}. The kind
of cavities involved in \eref{ec:generalfunctional} are
\textsl{\mbox{1-particle}
cavities} because they are such that 
if a particle of species $i$ occupies a node of
$\mathcal{C}^{(k)}_i(\bi{s})$, it excludes all nodes of
$\mathcal{C}^{(k)}_j(\bi{s})$ to particles of species 
$j$, for any $j=1,\dots,p$ (including $i$)\footnote{In
previous works this kind of object has been referred to as ``0D cavity'',
but as we will redefine this concept later on, this new nomenclature
is preferred.}.

Associated to cavities are \textsl{0D density profiles.}
If $\boldsymbol{\mathcal{C}}=\bigl(\mathcal{C}_1,\ldots,
\mathcal{C}_p\bigr)$ denotes a cavity, a 0D density profile associated
to $\boldsymbol{\mathcal{C}}$ is a density vector, denoted
$\brho_{\boldsymbol{\mathcal{C}}}(\bi{s})=
\bigl(\rho_{\mathcal{C}_1}(\bi{s}),
\ldots,\rho_{\mathcal{C}_p}(\bi{s})\bigr)$, with support 
$\boldsymbol{\mathcal{C}}$ (i.e.\
$\rho_{\mathcal{C}_i}$
has support $\mathcal{C}_i$). Note that for such 
$\brho_{\boldsymbol{\mathcal{C}}}(\bi{s})$, the
corresponding weighted density $\sum_{i=1}^p
\sum_{\bi{t}\in\mathcal{C}_i}
\rho_{\mathcal{C}_i}(\bi{t})$ is the average
occupancy of the lattice subset defined
by $\bigcup_{i=1}^p\mathcal{C}^{(k)}_i(\bi{s})$.

Given a specific model, the functional in \eref{ec:generalfunctional}
will be completely determined by the coefficients
$\{a_k\}$ and the family of subsets $\{\mathcal{C}^{(k)}_i(\bi{s})\}$.
As it was shown in a previous work (Lafuente and Cuesta~2004),
these unknowns can be uniquely determined by just imposing that
the approximate functional \eref{ec:generalfunctional}
recovers the exact limit for \textit{any} 0D density profile
of any \mbox{1-particle} cavity. This condition can be expressed as
\begin{equation}
\label{ec:0dcondition}
\mathcal{F}^{\mathrm{ex}}_\mathrm{FM}[
\brho_{\boldsymbol{\mathcal{C}}}]=\Phi_0(\eta)
\mbox{ for \textsl{any} \mbox{1-particle} cavity $\boldsymbol{\mathcal{C}}$},
\end{equation}
with $\eta=\sum_{i=1}^p\sum_{\bi{t}\in
\mathcal{C}_i} \rho_{\mathcal{C}_i}(\bi{t})$ 
the average occupancy of the cavity $\boldsymbol{\mathcal{C}}$.
From this, it is easy to notice that if the functional
$\mathcal{F}^{\mathrm{ex}}_\mathrm{FM}[\brho]$
satisfies condition \eref{ec:0dcondition} for the 0D density
profile $\brho_{\boldsymbol{\mathcal{C}}}(\bi{s})$,
then this condition immediately holds for \textit{any} 0D density
profile $\brho_{\boldsymbol{\mathcal{C}}'}(\bi{s})$
such that $\boldsymbol{\mathcal{C}}'\subset\boldsymbol{\mathcal{C}}$
(inclusion here must be understood componentwise), simply because 
$\brho_{\boldsymbol{\mathcal{C}}'}$ is nothing but a
particular choice of $\brho_{\boldsymbol{\mathcal{C}}}$.
Therefore, we can focus on the set of \textsl{maximal cavities}
(Lafuente and Cuesta~2002), which are cavities
not contained in any other cavity. Thus condition \eref{ec:0dcondition}
holds if and only if it holds for the set of maximal cavities,
i.e.\ we can replace \eref{ec:0dcondition} by
\begin{equation}
\label{ec:0dconditionmaximal}
\mathcal{F}^{\mathrm{ex}}_\mathrm{FM}[
\brho_{\boldsymbol{\mathcal{C}}}]=\Phi_0(\eta)
\mbox{ for \textsl{any maximal} \mbox{1-particle} cavity
$\boldsymbol{\mathcal{C}}$}.
\end{equation}
In (Lafuente and Cuesta~2004) we showed (and this will be proved
in full detail in this paper) that once we have determined the
set of maximal cavities ---which only depends on the geometry of
the interaction--- condition \eref{ec:0dconditionmaximal}
completely determines the functional~\eref{ec:generalfunctional}
for the given system. In other words, this condition uniquely fixes
both sets $\{a_k\}$ and $\{\mathcal{C}^{(k)}_i(\bi{s})\}$.

As an illustration of the procedure,
we refer the reader to (Lafuente and Cuesta~2004) for some
specific examples of one-component models. Here, in order to review the
recipe and to extend
the collection of examples, we will apply the procedure
to a one-dimensional binary hard-rod mixture.

Let us consider a system in the one-dimensional lattice $\mathbb{Z}$ with
two species of hard rods: the largest of length $\sigma_\mathrm{L}=3$
and the smallest $\sigma_\mathrm{S}=2$ (both in lattice
spacing units). This model represents one example of the
non-additive case which was exactly solved in (Lafuente and Cuesta~2002).

The first step of the procedure amounts to determining
the set of maximal \mbox{1-particle} cavities of the system. As in previous
examples, we will use a diagrammatic notation for cavities
$\boldsymbol{\mathcal{C}}=\bigl(\mathcal{C}_\mathrm{L},
\mathcal{C}_\mathrm{S}\bigr)$.
Remember that each $\mathcal{C}_i$ ($i=\mathrm{L},\mathrm{S}$)
is a subset of lattice nodes and thus we can associate 
$\mathcal{C}_i$ to the labeled graph whose vertices are the lattice
nodes in $\mathcal{C}_i\subset\mathcal{L}$ and whose edges
are the bonds linking nearest neighbors in the embedding lattice
$\mathcal{L}$. In this representation we could have, for instance,
\begin{equation}
\fl
\boldsymbol{\mathcal{C}}=\bigl(\{s,s+1,s+2\},\{s,s+1\}\bigr)=
\bigl(\dntriplelabel{$s$}{}{},\dnhlabel{$s$}{}\bigr)=
\dnllabel{$s$}{}{} \qquad (s\in\mathbb{Z}), 
\end{equation}
where we have merged the two components of
$\boldsymbol{\mathcal{C}}$ in the last diagram using different
colors to denote each species (white for the
large and black for the small)\footnote{Notice that all nodes
of the graph carry their corresponding label; the fact
that we write a single label $s$ on one of the nodes
of the graph simply aims at avoiding clumsy notations.}.
With this notation, the set of maximal 1-particle cavities for the
hard-rod binary mixture is
\begin{equation}
\label{ec:maximalset}
\mathcal{W}_\mathrm{max}=\{\dnllabel{$s$}{}{},
\dnrlabel{$s$}{}{}\,|\,s\in\mathbb{Z}\}.
\end{equation}
Since we want the functional in \eref{ec:generalfunctional}
to recover the exact limit for all 0D density profiles
with support any cavity in $\mathcal{W}_\mathrm{max}$,
there must be in \eref{ec:generalfunctional} a contribution
for each $\boldsymbol{\mathcal{C}}\in\mathcal{W}_\mathrm{max}$. Notice
that the weighted density $n^{(k)}(\bi{s})$ can be
identified with the associated cavity
$\boldsymbol{\mathcal{C}}^{(k)}(\bi{s})=\bigl(\mathcal{C}^{(k)}_1(\bi{s}),
\ldots,\mathcal{C}^{(k)}_p(\bi{s})\bigr)$
and the latter with the corresponding diagram. Therefore,
we can write this contribution to \eref{ec:generalfunctional} as
\begin{equation}
\sum_{s\in\mathbb{Z}}\bigl[
a(\dnl) \Phi_0(\dnllabel{$s$}{}{})+a(\dnr)
\Phi_0(\dnrlabel{$s$}{}{}) \bigr].
\end{equation}
Furthermore, if we want the
functional to yield the exact 0D limit, coefficients
$a(\dnl)$ and $a(\dnr)$ must be equal to $1$ (this is always
true for the coefficients associated to maximal cavities).
This first analysis provides us with an initial guess for the functional
of the system, namely
\begin{equation}
\label{ec:guess1}
\mathcal{F}^\mathrm{ex}_\mathrm{FM1}[\brho]=\sum_{s\in\mathbb{Z}}
\bigl[\Phi_0(\dnllabel{$s$}{}{})+\Phi_0(\dnrlabel{$s$}{}{})\bigr]. 
\end{equation}
If this were the final functional, then it should satisfy
condition \eref{ec:0dconditionmaximal}. If we take, for instance,
$\brho_{\boldsymbol{\mathcal{C}}}$, with
$\boldsymbol{\mathcal{C}}=\dnllabel{}{$t$}{}$, as a test 0D density 
profile and evaluate the functional $\mathcal{F}^{\mathrm{ex}}_\mathrm{FM1}$,
we obtain
\begin{eqnarray}
\label{ec:contrib1}
\fl
\mathcal{F}^{\mathrm{ex}}_\mathrm{FM1}[\brho_{\dnllabel{}{$t$}{}}]=
\Phi_0\bigl(\dnllabel{}{$t$}{}\bigr)+\Phi_0\bigl(\dnclabel{}{$t$}{}\bigr)+
\Phi_0\bigl(\dnhllabel{}{$t$}\bigr)+\Phi_0\bigl(\dnhllabel{$t$}{}\bigr)\nonumber \\
+\Phi_0\bigl(\dndoblelabel{}{$t$}\bigr)+\Phi_0\bigl(\dnhlabel{$t$}{}\bigr)+
\Phi_0\bigl(\dnnodolabel{$t-1$}\,\quad\bigr)+
\Phi_0\bigl(\dnplabel{$t-1$}\,\quad\bigr)+
2\Phi_0\bigl(\dnplabel{$t+1$}\,\quad\bigr),
\end{eqnarray}
where we have used that $\Phi_0(0)=0$ and that
$\brho_{\boldsymbol{\mathcal{C}}}$
is zero outside $\boldsymbol{\mathcal{C}}=\dnllabel{}{$t$}{}$
(which implies, for instance, that $\Phi_0\bigl(\dnllabel{}{}{$t$}\bigr)=
\Phi_0\bigl(\dnhllabel{}{$t$}\bigr)$). Now, if we check \eref{ec:contrib1}
with condition \eref{ec:0dconditionmaximal}, we note that apart from
the exact contribution $\Phi_0\bigl(\dnllabel{}{$t$}{}\bigr)$ we also have
a few spurious terms. Therefore the initial guess \eref{ec:guess1} needs
to be modified in order to eliminate these terms. Taking into
account the general form \eref{ec:generalfunctional} and the
procedure explained in (Lafuente and Cuesta~2004), we propose
as a second guess
\begin{equation}
\label{ec:guess2}
\fl
\mathcal{F}^\mathrm{ex}_\mathrm{FM2}[\brho]=
\sum_{s\in\mathbb{Z}} \bigl[
\Phi_0(\dnllabel{$s$}{}{})+\Phi_0(\dnrlabel{$s$}{}{}) 
+a(\dnc)  \Phi_0\bigl(\dnclabel{}{$s$}{})\bigr],
\end{equation}
where the coefficient $a(\dnc)$ has to be determined. The only way
to remove the term $\Phi_0\bigl(\dnclabel{}{$t$}{}\bigr)$
in \eref{ec:contrib1} is to set $a(\dnc)=-1$; thus
\begin{equation}
\label{ec:contrib2}
\fl
\mathcal{F}^\mathrm{ex}_\mathrm{FM2}[\brho_{\dnllabel{}{$t$}{}}]=
\Phi_0\bigl(\dnllabel{}{$t$}{}\bigr)+
\Phi_0\bigl(\dndoblelabel{}{$t$}\bigr)+\Phi_0\bigl(\dnhllabel{$t$}{}\bigr)+
\Phi_0\bigl(\dnnodolabel{$t-1$}\,\quad\bigr)+
\Phi_0\bigl(\dnplabel{$t+1$}\,\quad\bigr).
\end{equation} 
We can now iterate the procedure to remove the term
$\Phi_0\bigl(\dndoblelabel{}{$t$}\bigr)$. The next guess is
\begin{equation}
\label{ec:guess3}
\fl
\mathcal{F}^\mathrm{ex}_\mathrm{FM3}[\brho]=
\sum_{s\in\mathbb{Z}} \bigl[
\Phi_0(\dnllabel{$s$}{}{})+\Phi_0(\dnrlabel{$s$}{}{}) 
-\Phi_0\bigl(\dnclabel{}{$s$}{})+a(\dndoble)
\Phi_0\bigl(\dndoblelabel{$s$}{}\bigr)\bigr].
\end{equation}
Choosing $a(\dndoble)=-1$ we get $\mathcal{F}^\mathrm{ex}_\mathrm{FM3}
[\brho_{\dnllabel{}{$t$}{}}]=
\Phi_0\bigl(\dnllabel{}{$t$}{}\bigr)$, and the exact limit is
recovered for any 0D density profile with support any
maximal \mbox{1-particle} cavity in the set
$\{\dnllabel{$s$}{}{}\,|\,s\in\mathbb{Z}\}\subset\mathcal{W}_\mathrm{max}$.
By symmetry, it is easy to verify that the functional
\eref{ec:guess3} needs no additional terms to fully satisfy
\eref{ec:0dconditionmaximal}; therefore the functional
\begin{equation}
\label{ec:finalfunctional}
\fl
\mathcal{F}^\mathrm{ex}_\mathrm{FM}[\brho]=
\sum_{s\in\mathbb{Z}}\bigl[
\Phi_0\bigl(\dnllabel{$s$}{}{}\bigr)+\Phi_0\bigl(\dnrlabel{$s$}{}{}\bigr)
-\Phi_0\bigl(\dnclabel{}{$s$}{}\bigr)-\Phi_0\bigl(\dndoblelabel{$s$}{}{}\bigr)
\bigr]
\end{equation}
is exact when evaluated at any 0D density profile.
Moreover, this functional coincides with the \textit{exact}
one, obtained by Lafuente and Cuesta (2002) through a different,
more involved method.

As it was discussed in (Lafuente and Cuesta~2004), there are two 
remarkable features in this procedure. First of all, if we start
off the iteration with the terms associated to maximal 
\mbox{1-particle} cavities, then all clusters defining the weighted densities 
$n^{(k)}(\bi{s})$ in \eref{ec:generalfunctional} are also
\mbox{1-particle} cavities of the system. Not only that, they are
\textsl{intersections of the maximal cavities}. Secondly, once we
adopt the first ansatz $\mathcal{F}^\mathrm{ex}_\mathrm{FM1}[\brho]$,
there is a unique functional which fulfills condition 
\eref{ec:0dconditionmaximal} (in
other words, the specific scheme one follows in order to
remove the spurious terms is irrelevant).
This last statement will be rigorously proved in the next section.

\section{A cluster expansion for the free-energy density functional}
\label{sec:sec3}
If we analyze the general expression \eref{ec:generalfunctional}
we notice that the FM excess free-energy functional is built
from \mbox{1-particle} cavity contributions. Furthermore, if we considered
the system defined only in one of these cavities, say
$\boldsymbol{\mathcal{C}}^{(k)}(\bi{s})=
\bigl(\mathcal{C}^{(k)}_1(\bi{s}),\ldots,
\mathcal{C}^{(k)}_p(\bi{s})\bigr)$, the exact
excess free energy would be $\Phi_0\bigl(n^{(k)}(\bi{s})\bigr)$.
Therefore, if $\mathcal{F}^{\mathrm{ex}}_{\boldsymbol{\mathcal{C}}}[\brho]$
denotes the exact excess functional
of a system defined in cavity $\boldsymbol{\mathcal{C}}$,
we can rewrite \eref{ec:generalfunctional} as\footnote{This rewriting of
\eref{ec:generalfunctional} is particularly interesting because the
fact that any explicit reference to the function $\Phi_0(\eta)$ has disappeared
makes it applicable to cavities more general than the \mbox{1-particle} cavities
considered so far. More on this later.}
\begin{equation}
\label{ec:generalfunctional2}
\mathcal{F}^{\mathrm{ex}}_{\mathrm{FM}}[\brho]=
\sum_{\boldsymbol{\mathcal{C}}\mbox{\scriptsize{} cavity}}
a(\boldsymbol{\mathcal{C}})
\mathcal{F}^{\mathrm{ex}}_{\boldsymbol{\mathcal{C}}}[\brho].
\end{equation}

Using this cluster notation, the exact excess free-energy functional
of the system under consideration can be denoted
$\mathcal{F}^{\mathrm{ex}}_{\boldsymbol{\mathcal{L}}}[\brho]$,
where $\boldsymbol{\mathcal{L}}$ is the (multicomponent) cluster
$(\mathcal{L},\ldots,\mathcal{L})$. Now taking into account that, in 
general, functional \eref{ec:generalfunctional2}
is approximate, the exact one can be written
\begin{equation}
\label{ec:exactfunctional}
\mathcal{F}^{\mathrm{ex}}_{\boldsymbol{\mathcal{L}}}[\brho]\equiv
\Psi[\brho]+\sum_{\boldsymbol{\mathcal{C}}\mbox{\scriptsize{} 
cavity}}a(\boldsymbol{\mathcal{C}})
\mathcal{F}^{\mathrm{ex}}_{\boldsymbol{\mathcal{C}}}[\brho],
\end{equation}
where $\Psi[\brho]$ is the error of the FM approximation (obviously
an unknown functional).

As mentioned in the previous section, the coefficients
$a(\boldsymbol{\mathcal{C}})$ are uniquely determined by condition
\eref{ec:0dconditionmaximal}, which, as we have shown, can be implemented
through an iterative procedure determined
by the inclusion relations of the intersections of the cavities involved
(considered as subsets of lattice nodes).
In this section, we will show that expression \eref{ec:exactfunctional}
is a particular case of \textsl{M\"obius inversion formula}, 
a major result of the \textsl{theory of partially ordered sets}.
The reader is strongly advised to consult the specialized
literature on this subject (Rota~1964, Aigner~1979, Stanley~1999).
Here, we will just introduce the necessary mathematical
background to provide a comprehensible and rigorous foundation to LFMT. 

\subsection{M\"obius inversion formula in a nutshell}
A \textsl{partially ordered set} or \textsl{poset} is a set
together with a partial order relation denoted $\leq$
i.e.\ a binary relation satisfying reflexivity, antisymmetry
and transitivity). An example of poset is the set of clusters involved
in functional \eref{ec:finalfunctional}:
\begin{equation*}
\mathcal{W}\equiv\{\dnllabel{$s$}{}{},\dnrlabel{$s$}{}{},
\dnclabel{}{$s$}{},\dndoblelabel{$s$}{}{}\,|\,s\in\mathbb{Z}\},
\end{equation*}
with the order defined by 
$\boldsymbol{\mathcal{C}}\leq\boldsymbol{\mathcal{C}}'$
if and only if $\boldsymbol{\mathcal{C}}\subset\boldsymbol{\mathcal{C}}'$.
In general the order is partial because there may be pairs of
elements which are not comparable. In our example neither 
$\dnllabel{$s$}{}{}\leq\dnclabel{}{$s$}{}$
nor $\dnclabel{}{$s$}{}\leq\dnllabel{$s$}{}{}$ are true,
while, for instance, $\dnclabel{$s$}{}{}\leq\dnllabel{$s$}{}{}$. 

Let $\mathcal{W}$ be a poset. For any $x,y\in \mathcal{W}$
such that $x\leq y$ we define the \textsl{interval}
$[x,y]\equiv\{z\in\mathcal{W}\,|\,x\leq z\leq y\}$.
A poset is said to be \textsl{locally finite} if all its
intervals are finite. The set of all intervals
of poset $\mathcal{W}$ will be denoted by $\mathrm{Int}(\mathcal{W})$.

Given a locally finite poset $\mathcal{W}$, the
\textsl{incidence algebra} $I(\mathcal{W},\mathbb{K})$ of
$\mathcal{W}$ over field $\mathbb{K}$ is the set
of all mappings $f\colon \mathrm{Int}(\mathcal{W}) \rightarrow \mathbb{K}$
(we will write $f(x,y)$ for $f([x,y])$)
with the usual vector space structure and the inner product
$(f*g)(x,y)\equiv\sum_{z\in[x,y]} f(x,z) g(z,y)$ ($f$ and $g$ being
elements of $I(\mathcal{W},\mathbb{K})$). Note that this product
is well-defined because being $\mathcal{W}$
locally finite the sum contains a finite number of terms.
It is easy to check that $I(\mathcal{W},\mathbb{K})$
is an associative algebra with (two-sided) identity $\delta(x,y)\equiv 1$
if $[x,y]=[x,x]=\{x\}$ and $0$ otherwise. Another useful
function of $I(\mathcal{W},\mathbb{K})$ is the \textsl{zeta function}
$\zeta(x,y)=1$ for all $[x,y]\in\mathrm{Int}(\mathcal{W})$.

An important result for incidence algebras is (the present statement
is a simplified version of Proposition~3.6.2 on p.~114 of (Stanley~1999)):
\begin{prop}
\label{prop:prop1}
Let $f$ be an element of $I(\mathcal{W},\mathbb{K})$; then $f$
has (two-sided) inverse (i.e.\ there exists $f^{-1}\in
I(\mathcal{W},\mathbb{K})$ such that $f*f^{-1} = f^{-1}*f =\delta$)
if and only if $f(x,x)\neq 0$ for all
$x\in\mathcal{W}$.
\end{prop}

It follows from this proposition that the zeta function $\zeta$
of a locally finite poset $\mathcal{W}$ is invertible; its inverse
is the \textsl{M\"obius function} of $\mathcal{W}$ and is
denoted by $\mu_\mathcal{W}$. This function can be obtained recursively
from the definition of the inverse:
\begin{rec}
\label{rec:rec1}
\textbf{($\bzeta\boldsymbol{*}\bmu_{\boldsymbol{\mathcal{W}}}\boldsymbol{=}\bdelta$)}
\begin{equation}
\label{ec:recursion1}
\eqalign{\mu_\mathcal{W}(x,x)&=1,\quad\mbox{for all $x\in\mathcal{W}$,}\\
\mu_\mathcal{W}(x,y)&=-\sum_{x< 
\underset{\bbdot}{z} \leq y} \mu_{\mathcal{W}}(z,y),\quad
\mbox{for $x<y$ with $x,y\in\mathcal{W}$.}
}
\end{equation}
\end{rec}

\begin{rec}
\label{rec:rec2}
\textbf{($\bmu_{\boldsymbol{\mathcal{W}}}\boldsymbol{*}\bzeta \boldsymbol{=}\bdelta$)}
\begin{equation}
\label{ec:recursion2}
\eqalign{\mu_\mathcal{W}(x,x)&=1,\quad\mbox{for all $x\in\mathcal{W}$,}\\
\mu_\mathcal{W}(x,y)&=-\sum_{x\leq 
\underset{\bbdot}{z} < y} \mu_{\mathcal{W}}(x,z),\quad
\mbox{for $x<y$ with $x,y\in\mathcal{W}$.}
}
\end{equation}
\end{rec}
In both Recursion~1 and 2, we have used a dot to indicate
the summation variable.

We are now ready to formulate the key theorem
for the rigorous foundation of LFMT (from Stanley (1999),
Proposition~3.7.1 on p.~116):
\begin{theo}\textbf{\textsf{(M\"obius inversion formula)}}
\label{theo:theo1}
Let $\mathcal{W}$ be a poset such that for every $x\in\mathcal{W}$
the subset $\{y\in\mathcal{W}\,|\,y\leq x\}$
is finite. Let $f,g\colon
\mathcal{W}\longrightarrow V(\mathbb{K})$, ($V(\mathbb{K})$
being a vector space over field $\mathbb{K}$). Then
\begin{equation}
\label{ec:mobius1}
f(x)=\sum_{\underset{\bbdot}{y}\leq x} g(y)\quad\mbox{for all $x\in\mathcal{W}$}
\end{equation}
if and only if
\begin{equation}
\label{ec:mobius2}
g(x)=\sum_{\underset{\bbdot}{y}\leq x} f(y)\mu_\mathcal{W}(y,x)\quad
\mbox{for all $x\in\mathcal{W}$}.
\end{equation}
\end{theo}

In the following, we will show that M\"obius inversion
formula can be applied to obtain a cluster expansion
of the exact free-energy functional of a general lattice gas
with arbitrary interaction. After that, it will be straightforward
to prove that LFMT amounts to taking a particular truncation
of this expansion.

\subsection{Cluster expansion of the free-energy functional}
The cluster expansion we will propose here is based on
the formulation of the cluster variation method by Morita~(1994).
There, M\"obius inversion is used to approximate
the entropy of a lattice model by a linear combination
of the exact entropies of the same model restricted to
a family of lattice clusters. This is exactly the same
idea we find in approximation \eref{ec:generalfunctional2},
but here it is applied to the free-energy density functional.

Let us consider a multicomponent lattice gas with underlying
lattice $\mathcal{L}$. Let us assume that we have a poset
$\mathcal{W}$ whose elements are (multicomponent)
clusters of lattice $\mathcal{L}$ (as in the previous example),
such that cluster $\boldsymbol{\mathcal{L}}$ is in $\mathcal{W}$.
Note that by definition $\boldsymbol{\mathcal{C}}\leq\boldsymbol{\mathcal{L}}$
for all $\boldsymbol{\mathcal{C}}\in\mathcal{W}$. Now, let us consider the mapping
$f(\boldsymbol{\mathcal{C}})\equiv\mathcal{F}_{\boldsymbol{\mathcal{C}}}[\brho]$,
where $\boldsymbol{\mathcal{C}}\in\mathcal{W}$ and
$\mathcal{F}_{\boldsymbol{\mathcal{C}}}[\brho]$ is the
\textit{exact} free-energy functional of the given
model restricted to cluster $\boldsymbol{\mathcal{C}}$.
Particularizing $x=\boldsymbol{\mathcal{L}}$ in \eref{ec:mobius2}
and taking into account that $\mu_\mathcal{W}(
\boldsymbol{\mathcal{L}},\boldsymbol{\mathcal{L}})=1$,
we have for the exact free-energy functional of the system
\begin{equation}
\label{ec:clusterexpansion}
\mathcal{F}_{\boldsymbol{\mathcal{L}}}[\brho]=
\Psi_{\boldsymbol{\mathcal{L}}}[\brho]
+\sum_{\boldsymbol{\mathcal{C}}<\boldsymbol{\mathcal{L}}}
[-\mu_\mathcal{W}(\boldsymbol{\mathcal{C}},\boldsymbol{\mathcal{L}})]
\mathcal{F}_{\boldsymbol{\mathcal{C}}}[\brho],
\end{equation}
where $\Psi_{\boldsymbol{\mathcal{L}}}[\brho]=g(\boldsymbol{\mathcal{L}})$ is an
unknown functional. Expression \eref{ec:clusterexpansion} provides
a cluster expansion of the \textsl{exact} total free-energy density functional
of an \textsl{arbitrary} system. In general, functional
$\Psi_{\boldsymbol{\mathcal{L}}}[\brho]$ cannot be
computed exactly, but for suitable choices of the cluster set $\mathcal{W}$ some
of its properties can be derived.

Let us make a particular choice: let us assume that $\mathcal{W}$ consists
of $\boldsymbol{\mathcal{L}}$ as well as every non-empty intersection
of the clusters of certain set $\mathcal{W}_\mathrm{max}$.
We will then show that $\Psi_{\boldsymbol{\mathcal{L}}}[\brho]$ vanishes
for every density profile with support \textsl{any} cluster in 
$\mathcal{W}_\mathrm{max}$. (Note that, by construction, the clusters 
in $\mathcal{W}_\mathrm{max}$ are the maximal elements of 
$\mathcal{W}-\{\boldsymbol{\mathcal{L}}\}$ with respect to the
order relation.)

Let $\boldsymbol{\mathcal{D}}\in\mathcal{W}_\mathrm{max}$ and
$\brho_{\boldsymbol{\mathcal{D}}}$ a density profile with
support $\boldsymbol{\mathcal{D}}$. If $\boldsymbol{\mathcal{C}}\in\mathcal{W}$
then we have
\begin{equation*}
\mathcal{F}_{\boldsymbol{\mathcal{C}}}[\brho_{\boldsymbol{\mathcal{D}}}]=
\mathcal{F}_{\boldsymbol{\mathcal{C}}
\cap\boldsymbol{\mathcal{D}}}[\brho_{\boldsymbol{\mathcal{D}}}],
\end{equation*}
with $\mathcal{F}_\varnothing[\brho_{\boldsymbol{\mathcal{D}}}]=0$.
By construction, if $\boldsymbol{\mathcal{C}}\cap
\boldsymbol{\mathcal{D}}\neq\varnothing$ then
$\boldsymbol{\mathcal{C}}\cap\boldsymbol{\mathcal{D}}\in\mathcal{W}$.
Taking into account that $\mathcal{F}_{\boldsymbol{\mathcal{L}}}
[\brho_{\boldsymbol{\mathcal{D}}}]=
\mathcal{F}_{\boldsymbol{\mathcal{D}}}[\brho_{\boldsymbol{\mathcal{D}}}]$,
from \eref{ec:clusterexpansion} we obtain
\begin{equation}
\fl
\Psi_{\boldsymbol{\mathcal{L}}}[\brho_{\boldsymbol{\mathcal{D}}}]=
\sum_{\boldsymbol{\mathcal{C}}\in\mathcal{W}}
\mu_\mathcal{W}(\boldsymbol{\mathcal{C}},\boldsymbol{\mathcal{L}})
\mathcal{F}_{\boldsymbol{\mathcal{C}}\cap\boldsymbol{\mathcal{D}}}
[\brho_{\boldsymbol{\mathcal{D}}}]
=\sum_{\underset{\bbdot}{\boldsymbol{\mathcal{E}}}\leq \boldsymbol{\mathcal{D}}}
\mathcal{F}_{\boldsymbol{\mathcal{E}}}[\brho_{\boldsymbol{\mathcal{D}}}]
\sum_{\underset{\bbdot}{\boldsymbol{\mathcal{C}}}\cap\boldsymbol{\mathcal{D}}=
\boldsymbol{\mathcal{E}}}\mu_\mathcal{W}(\boldsymbol{\mathcal{C}},\boldsymbol{\mathcal{L}}).
\end{equation}
Now, we can use corollary~\ref{cor:cor1} in~\ref{sec:app} to show that
\begin{equation}
\label{ec:0drecursion}
\sum_{\underset{\bbdot}{\boldsymbol{\mathcal{C}}}\cap\boldsymbol{\mathcal{D}}=
\boldsymbol{\mathcal{E}}}\mu_\mathcal{W}(\boldsymbol{\mathcal{C}},
\boldsymbol{\mathcal{L}})=0
\mbox{ for all $\boldsymbol{\mathcal{D}}\in\mathcal{W}_\mathrm{max}$ and
$\boldsymbol{\mathcal{E}}<\boldsymbol{\mathcal{L}}$.}
\end{equation}
Therefore, the unknown functional $\Psi_{\boldsymbol{\mathcal{L}}}$ fulfills
the condition
\begin{equation}
\label{ec:reduccion0d}
\Psi_{\boldsymbol{\mathcal{L}}}[\brho_{\boldsymbol{\mathcal{C}}}]=0
\mbox{ for all $\boldsymbol{\mathcal{C}}\in\mathcal{W}-\{\boldsymbol{\mathcal{L}}\}$},
\end{equation}
since for every such $\boldsymbol{\mathcal{C}}$ there exists at least one
$\boldsymbol{\mathcal{D}}\in\mathcal{W}_\mathrm{max}$
such that $\boldsymbol{\mathcal{C}}\leq\boldsymbol{\mathcal{D}}$,
and so $\brho_{\boldsymbol{\mathcal{C}}}$ is a particular
case of $\brho_{\boldsymbol{\mathcal{D}}}$.

Now, if we approximate the exact free-energy functional
in \eref{ec:clusterexpansion} by the truncation
\begin{equation}
\label{ec:appfunctional}
\mathcal{F}_\mathrm{app}[\brho]=\sum_{\boldsymbol{\mathcal{C}}<
\boldsymbol{\mathcal{L}}}\bigl[-\mu_\mathcal{W}(\boldsymbol{\mathcal{C}},
\boldsymbol{\mathcal{L}})\bigr]\mathcal{F}_{\boldsymbol{\mathcal{C}}}[\brho],
\end{equation}
the previous result \eref{ec:reduccion0d} guarantees that this approximation
is exact for any density profile with support any
cluster of $\mathcal{W}-\{\boldsymbol{\mathcal{L}}\}$.

At this point, we have shown that given a specific lattice model
there exists an approximation of the free-energy functional,
given by \eref{ec:appfunctional}, which is exact when
the system is restricted to certain set of cavities
(namely, those in $\mathcal{W}-\{\boldsymbol{\mathcal{L}}\}$).
But, is it unique? Or in other words, how many free-energy
functionals of the form
\begin{equation}
\label{ec:generalappfunctional}
\mathcal{F}_\mathrm{app}[\brho]=\sum_{\boldsymbol{\mathcal{C}}<
\boldsymbol{\mathcal{L}}} a(\boldsymbol{\mathcal{C}})
\mathcal{F}_{\boldsymbol{\mathcal{C}}}[\brho]
\end{equation}
are there which are exact in $\mathcal{W}-\{\boldsymbol{\mathcal{L}}\}$?
To end this section, we will prove that the only one with such
property is \eref{ec:appfunctional}. Before going
into the technical details, notice that, as in the proof
of \eref{ec:reduccion0d}, it suffices that the above condition
holds for the clusters in $\mathcal{W}_\mathrm{max}$
(the set of maximal elements of $\mathcal{W}-\{\boldsymbol{\mathcal{L}}\}$).

Let us suppose that we have an approximation like
\eref{ec:generalappfunctional} such that it satisfies
\begin{equation}
\label{ec:app0dcondition}
\mathcal{F}_\mathrm{app}[\brho_{\boldsymbol{\mathcal{D}}}]=
\mathcal{F}_{\boldsymbol{\mathcal{L}}}
[\brho_{\boldsymbol{\mathcal{D}}}]
\end{equation}
for every cluster $\boldsymbol{\mathcal{D}}\in\mathcal{W}_\mathrm{max}$.
Since $\mathcal{F}_{\boldsymbol{\mathcal{C}}}[\brho_{\boldsymbol{\mathcal{D}}}]=
\mathcal{F}_{\boldsymbol{\mathcal{C}}\cap\boldsymbol{\mathcal{D}}}
[\brho_{\boldsymbol{\mathcal{D}}}]$, we have from \eref{ec:generalappfunctional}
and \eref{ec:app0dcondition}
\begin{equation}
\label{ec:app0dcondition2}
\fl
\mathcal{F}_\mathrm{app}[\brho_{\boldsymbol{\mathcal{D}}}]=
\mathcal{F}_{\boldsymbol{\mathcal{D}}}
[\brho_{\boldsymbol{\mathcal{D}}}]=
a(\boldsymbol{\mathcal{D}})\mathcal{F}_{\boldsymbol{\mathcal{D}}}
[\brho_{\boldsymbol{\mathcal{D}}}]+ 
\sum_{\underset{\bbdot}{\boldsymbol{\mathcal{E}}}<
\boldsymbol{\mathcal{D}}}\mathcal{F}_{\boldsymbol{\mathcal{E}}}
[\brho_{\boldsymbol{\mathcal{D}}}]
\sum_{\underset{\bbdot}{\boldsymbol{\mathcal{C}}}\cap
\boldsymbol{\mathcal{D}}=\boldsymbol{\mathcal{E}}}
 a(\boldsymbol{\mathcal{C}}).
\end{equation}
This condition should be satisfied regardless
the functional form of $\mathcal{F}_{\boldsymbol{\mathcal{C}}}[\brho]$,
therefore \eref{ec:app0dcondition2} can be cast in the recursion
\begin{equation}
\label{ec:recur1}
\eqalign{
a(\boldsymbol{\mathcal{D}})=1\mbox{ for all $\boldsymbol{\mathcal{D}}\in
\mathcal{W}_\mathrm{max}$},\\
\sum_{\underset{\bbdot}{\boldsymbol{\mathcal{C}}}\cap\boldsymbol{\mathcal{D}}
=\boldsymbol{\mathcal{E}}} a(\boldsymbol{\mathcal{C}})=0
\mbox{ for all $\boldsymbol{\mathcal{E}}<\boldsymbol{\mathcal{D}}$.}}
\end{equation}
By defining $a(\boldsymbol{\mathcal{L}})\equiv -1$, the first of these
two equations can be rewritten as
$\sum_{\underset{\bbdot}{\boldsymbol{\mathcal{C}}}\cap\boldsymbol{\mathcal{D}}
=\boldsymbol{\mathcal{D}}} a(\boldsymbol{\mathcal{C}})=0$,
so both equations can be gathered in the single one
\begin{equation}
\label{ec:recur2}
\sum_{\underset{\bbdot}{\boldsymbol{\mathcal{C}}}\cap\boldsymbol{\mathcal{D}}
=\boldsymbol{\mathcal{E}}} a(\boldsymbol{\mathcal{C}})=0
\mbox{ for all $\boldsymbol{\mathcal{D}}\in\mathcal{W}_\mathrm{max}$
and $\boldsymbol{\mathcal{E}}<\boldsymbol{\mathcal{L}}$.}
\end{equation}
From \eref{ec:recur1} it is clear that this recursion has a unique
solution. On the other hand, \eref{ec:recur2} shows that it is formally
identical to \eref{ec:0drecursion}. Therefore, since it is linear,
the solution must be $a(\boldsymbol{\mathcal{C}})=\lambda
\mu_\mathcal{W}(\boldsymbol{\mathcal{C}},\boldsymbol{\mathcal{L}})$
for some constant $\lambda$. Choosing $\boldsymbol{\mathcal{C}}=
\boldsymbol{\mathcal{L}}$ shows that $\lambda=-1$; thus
\begin{equation}
a(\boldsymbol{\mathcal{C}})=-\mu_\mathcal{W}(\boldsymbol{\mathcal{C}},
\boldsymbol{\mathcal{L}})\mbox{ for all $\boldsymbol{\mathcal{C}}
\in\mathcal{W}$}.
\end{equation} 
In other words, \textsl{the only functional of type
\eref{ec:generalappfunctional} which is exact when restricted to the 
set of clusters $\mathcal{W}-\{\boldsymbol{\mathcal{L}}\}$ is
\eref{ec:appfunctional}}.

Let us summarize the main results we have obtained in the following
\begin{theo}
\label{theo:theo2}
Given an arbitrary lattice model, a certain set of clusters
$\mathcal{W}_\mathrm{max}$ and the poset $\mathcal{W}$
formed by all non-empty intersections of elements of $\mathcal{W}_\mathrm{max}$
as well as the cluster $\boldsymbol{\mathcal{L}}$, then there
exists a unique functional of type \eref{ec:generalappfunctional}
which is exact when evaluated at density profiles
$\brho_{\boldsymbol{\mathcal{C}}}$
with support any cluster $\boldsymbol{\mathcal{C}}\in\mathcal{W}-
\{\boldsymbol{\mathcal{L}}\}$. This functional is given by
\begin{equation}
\label{ec:finalappfunctional}
\mathcal{F}_\mathrm{app}[\brho]=\sum_{\boldsymbol{\mathcal{C}}
\in\mathcal{W}-\{\boldsymbol{\mathcal{L}}\}}
\bigl[-\mu_\mathcal{W}(\boldsymbol{\mathcal{C}},
\boldsymbol{\mathcal{L}})\bigr] \mathcal{F}_{\boldsymbol{\mathcal{C}}}[\brho],
\end{equation}
where the integer coefficients
$\mu_\mathcal{W}(\boldsymbol{\mathcal{C}},\boldsymbol{\mathcal{L}})$
are defined by either recursion~\ref{rec:rec1} (ec.~\eref{ec:recursion1})
or recursion~\ref{rec:rec2} (ec.~\eref{ec:recursion2}).
\end{theo}

For some special cases, formula \eref{ec:finalappfunctional} is
exact. One of them occurs if $\mathcal{F}_{\boldsymbol{\mathcal{C}}}[\brho]$
is a local functional of $\brho(\bi{s})$, i.e.
\begin{equation}
\mathcal{F}_{\boldsymbol{\mathcal{C}}}[\brho]=\sum_{i=1}^p
\sum_{\bi{s}\in\mathcal{C}_i}\phi\bigl(\rho_i(\bi{s})\bigl).
\end{equation}
With this particular choice for
$\mathcal{F}_{\boldsymbol{\mathcal{C}}}[\brho]$,
the ``error'' functional \eref{ec:clusterexpansion} becomes
\begin{equation*}
\Psi_{\boldsymbol{\mathcal{L}}}[\brho]=
\sum_{\boldsymbol{\mathcal{C}}\in\mathcal{W}}
\mu_\mathcal{W}(\boldsymbol{\mathcal{C}},\boldsymbol{\mathcal{L}})
\mathcal{F}_{\boldsymbol{\mathcal{C}}}[\brho]
=\sum_{i=1}^p\sum_{\bi{s}\in\mathcal{L}}
\phi\bigl(\rho_i(\bi{s})\bigl)h_i(\bi{s}),
\end{equation*}
where
$h_i(\bi{s})=
\sum_{\boldsymbol{\mathcal{C}}\in\mathcal{W}}
\mu_\mathcal{W}(\boldsymbol{\mathcal{C}},\boldsymbol{\mathcal{L}})
\chi_{\mathcal{C}_i}(\bi{s})$,
and $\chi_{\mathcal{C}_i}(\bi{s})=1$ if $\bi{s}\in\mathcal{C}_i$
and $0$ otherwise is the indicator function of
the set $\mathcal{C}_i$. If we now define the cluster
$\bsigma_i(\bi{s})\equiv(\varnothing,\dots,\varnothing,
\overset{(i)}{\dnplabel{$\bi{s}$}},\varnothing,\dots,\varnothing)$,
the function $h_i(\bi{s})$ can be rewritten as
\begin{equation*}
h_i(\bi{s})=
\sum_{\underset{\bbdot}{\boldsymbol{\mathcal{C}}}
\cap\bsigma_i(\bi{s})=\bsigma_i(\bi{s})}
\mu_\mathcal{W}(\boldsymbol{\mathcal{C}},\boldsymbol{\mathcal{L}}),
\end{equation*}
and because of corollary~\ref{cor:cor1} of~\ref{sec:app},
$h_i(\bi{s})=0$ for all $\bi{s}\in\mathcal{L}$ and
$i=1,\dots,p$. Hence $\Psi_{\boldsymbol{\mathcal{L}}}[\brho]=0$.

\section{Lattice fundamental measure theory revisited}
\label{sec:sec4}
The M\"obius formalism developed in section~\ref{sec:sec3} endows
us a very powerful alternative procedure to obtain
the FM excess free-energy functional for any lattice system.
The approximation \eref{ec:finalappfunctional} expresses the
free-energy functional as a truncated cluster expansion 
of the form \eref{ec:generalfunctional2}. Given that the ideal
part of the functional is local in $\brho(\bi{s})$ and, as we
have just shown, for local functionals this cluster expansion
is exact (regardless the choice of $\mathcal{W}_\mathrm{max}$),
the approximation is actually made on the excess free-energy
functional. If we now take for $\mathcal{W}_\mathrm{max}$ the
set of maximal \mbox{1-particle} cavities of a given lattice model,
then theorem~\ref{theo:theo2} tells us that the FM excess
free-energy functional will be given by
\begin{equation}
\label{ec:generalfunctional3}
\mathcal{F}^\mathrm{ex}_\mathrm{FM}[\brho]=\sum_{\boldsymbol{\mathcal{C}}\in
\mathcal{W}-\{\boldsymbol{\mathcal{L}}\}}
\bigr[-\mu_\mathcal{W}(\boldsymbol{\mathcal{C}},\boldsymbol{\mathcal{L}})\bigr]
\Phi_0\bigl(n_{\boldsymbol{\mathcal{C}}}[\brho]\bigr),
\end{equation}
where $n_{\boldsymbol{\mathcal{C}}}[\brho] \equiv
\sum_{i=1}^{p}\sum_{\bi{t}\in\mathcal{C}_i}\rho_i(\bi{t})$.
This expression is identical to \eref{ec:generalfunctional}, because
in $\mathcal{W}-\{\boldsymbol{\mathcal{L}}\}$
there will be clusters of different
shapes (labeled by $k\in\mathcal{I}$ in \eref{ec:generalfunctional})
and all their translates (labeled by $\bi{s}\in\mathcal{L}$ in
\eref{ec:generalfunctional}). But now, thanks to theorem~\ref{theo:theo2},
we know that this is the only functional one can obtain with 
this particular choice for $\mathcal{W}_\mathrm{max}$ which is
exact when evaluated at any 0D density profile with support any cluster
in $\mathcal{W}-\{\boldsymbol{\mathcal{L}}\}$.

A remark is in order here. If $\boldsymbol{\mathcal{C}}$
is a cluster of $\mathcal{W}-\{\boldsymbol{\mathcal{L}}\}$, and
$\boldsymbol{\mathcal{C}'}$ is a translation of $\boldsymbol{\mathcal{C}}$,
then $\mu_\mathcal{W}(\boldsymbol{\mathcal{C}},\boldsymbol{\mathcal{L}})=
\mu_\mathcal{W}(\boldsymbol{\mathcal{C}}',\boldsymbol{\mathcal{L}})$
because the M\"obius function $\mu_\mathcal{W}(x,y)$ only depends on the
interval $[x,y]$ (\textsl{c.f.} its definition in recursion~1) and
the translation operation is an obvious order-preserving isomorphism
between $[\boldsymbol{\mathcal{C}},\boldsymbol{\mathcal{L}}]$ and
$[\boldsymbol{\mathcal{C}}',\boldsymbol{\mathcal{L}}]$. This
justifies why the coefficients $a_k$ in \eref{ec:generalfunctional} are
independent of $\bi{s}\in\mathcal{L}$.

With this new formulation of LFMT it is a simple task to
obtain the unknowns in expression \eref{ec:generalfunctional}.
As an illustration, let us re-derive functional \eref{ec:finalfunctional}.
The first step is to fix the appropriate cluster poset $\mathcal{W}$,
which contains all non-empty intersections of maximal \mbox{1-particle}
cavities in the set $\mathcal{W}_\mathrm{max}$ described in
\eref{ec:maximalset}, as well as the cluster $\boldsymbol{\mathcal{L}}$.
Thus,
\begin{equation}
\fl
\mathcal{W}=\{\dnllabel{$s$}{}{},\dnrlabel{$s$}{}{},
\dnclabel{$s$}{}{},\dndoblelabel{$s$}{},\dnhllabel{$s$}{},\dnhrlabel{$s$}{},
\dnhlabel{$s$}{},\dnnodolabel{$s$},\dnplabel{$s$}\,|\,s\in\mathcal{L}\}
\cup\{\boldsymbol{\mathcal{L}}\}.
\end{equation}

\begin{figure}
\begin{center}
\includegraphics*[width=70mm, angle=0]{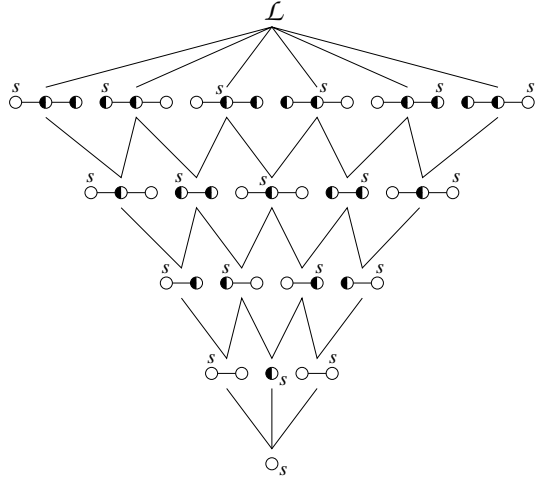}
\caption[]{\label{fig:fig1} The \textsl{Hasse diagram} of a finite poset
$\mathcal{W}$ is defined as the graph whose vertices are the elements
of the poset and there is an edge between elements $x$ and $y$ if
$x<y$ and there is no $z\in\mathcal{W}$ such that $x<z<y$. If $x<y$
then $y$ is drawn at higher level than $x$. This figure shows the
Hasse diagram of the interval $[\dnplabel{$s$},\boldsymbol{\mathcal{L}}]$.
Nodes connected through a descending path are ordered by transitivity.
Because of this reason, a Hasse diagram is a very practical way of
visualizing the order in a finite set.}
\end{center}
\end{figure}

The second and last step is to compute the M\"obius function
$\mu_\mathcal{W}(\boldsymbol{\mathcal{C}},\boldsymbol{\mathcal{L}})$ for every
$\boldsymbol{\mathcal{C}}\in\mathcal{W}-\{\boldsymbol{\mathcal{L}}\}$
(note that $\mu_\mathcal{W}(\boldsymbol{\mathcal{L}},\boldsymbol{\mathcal{L}})=1$).
This can be easily done by resorting to recursion~\ref{rec:rec1}. The
natural iteration prescribed by this recursion is to start with maximal
clusters (those in $\mathcal{W}_\mathrm{max}$)
and then to follow a decreasing path. Notice that in order
to obtain $\mu_\mathcal{W}(\boldsymbol{\mathcal{C}},\boldsymbol{\mathcal{L}})$
we only need the values of the M\"obius function for the clusters
in the interval $\bigl(\boldsymbol{\mathcal{C}},\boldsymbol{\mathcal{L}}\bigr]$.
To carry on this task, it is useful to draw the Hasse diagram
of the corresponding interval (see \fref{fig:fig1}), since it explicitly
shows the order structure of the latter.  

In our example, we should start with $\dnllabel{$s$}{}{}$ and
$\dnrlabel{$s$}{}{}$ (remember that the M\"obius function does not depend 
on $\bi{s}$, so what follows holds for any $\bi{s}\in\mathcal{L}$).
As both are maximal clusters, then
$\bigl(\dnllabel{$s$}{}{},\boldsymbol{\mathcal{L}}\bigr]=
\bigl(\dnrlabel{$s$}{}{},\boldsymbol{\mathcal{L}}\bigr]=
\{\boldsymbol{\mathcal{L}}\}$
(by definition, this is always true for maximal clusters).
By applying recursion~\ref{rec:rec1}, we obtain
\begin{equation*}
\mu_\mathcal{W}(\dnllabel{$s$}{}{},\boldsymbol{\mathcal{L}})=
\mu_\mathcal{W}(\dnrlabel{$s$}{}{},\boldsymbol{\mathcal{L}})=
-\mu_\mathcal{W}(\boldsymbol{\mathcal{L}},\boldsymbol{\mathcal{L}})=-1.
\end{equation*}
In decreasing order, the next set of clusters involves
$\dnclabel{$s$}{}{}$ and $\dndoblelabel{$s$}{}$, and we have 
\begin{equation*}
\fl
\bigl(\dnclabel{$s$}{}{},\boldsymbol{\mathcal{L}}\bigr]=
\{\dnllabel{$s$}{}{},\dnrlabel{$s$}{}{},\boldsymbol{\mathcal{L}}\},
\qquad
\bigl(\dndoblelabel{$s$}{},\boldsymbol{\mathcal{L}}\bigr]=
\{\dnllabel{$s$}{}{},\dnrlabel{}{$s$}{},\boldsymbol{\mathcal{L}}\}.
\end{equation*}
Thus $\mu_\mathcal{W}(\dnclabel{$s$}{}{},\boldsymbol{\mathcal{L}})=
\mu_\mathcal{W}(\dndoblelabel{$s$}{},\boldsymbol{\mathcal{L}})=1$.
Then we find $\dnhllabel{$s$}{}$ and $\dnhrlabel{$s$}{}$. By symmetry,
$\mu_\mathcal{W}(\dnhllabel{$s$}{},\boldsymbol{\mathcal{L}})=
\mu_\mathcal{W}(\dnhrlabel{$s$}{},\boldsymbol{\mathcal{L}})$, and
since
\begin{equation*}
\bigl(\dnhllabel{$s$}{},\boldsymbol{\mathcal{L}}\bigr]=
\{\dndoblelabel{$s$}{},\dnclabel{}{$s$}{},
\dnllabel{$s$}{}{},\dnllabel{}{$s$}{},\dnrlabel{}{$s$}{},
\boldsymbol{\mathcal{L}}\},
\end{equation*}
we will have $\mu_\mathcal{W}(\dnhllabel{$s$}{},\boldsymbol{\mathcal{L}})=0$.
Next we have $\dnhlabel{$s$}{}$ and $\dnnodolabel{$s$}$. The corresponding
intervals are, respectively,
\begin{equation*}
\fl
\bigl(\dnhlabel{$s$}{},\boldsymbol{\mathcal{L}}\bigr]=
\{\dnhllabel{$s$}{},\dnhrlabel{$s$}{},\dndoblelabel{$s$}{},
\dnclabel{$s$}{}{},\dnclabel{}{$s$}{},
\dnllabel{$s$}{}{},\dnllabel{}{$s$}{},\dnrlabel{$s$}{}{},\dnrlabel{}{$s$}{},
\boldsymbol{\mathcal{L}}\}
\end{equation*}
and
\begin{equation*}
\fl
\bigl(\dnnodolabel{$s$},\boldsymbol{\mathcal{L}}\bigr]=
\{\dnhllabel{$s$}{},\dnhrlabel{}{$s$},\dndoblelabel{$s$}{},
\dndoblelabel{}{$s$},\dnclabel{}{$s$}{},
\dnllabel{$s$}{}{},\dnllabel{}{$s$}{},\dnrlabel{}{$s$}{},
\dnrlabel{}{}{$s$},\boldsymbol{\mathcal{L}}\}.
\end{equation*}
Therefore $\mu_\mathcal{W}(\dnhlabel{$s$}{},\boldsymbol{\mathcal{L}})=0$
and $\mu_\mathcal{W}(\dnnodolabel{$s$},\boldsymbol{\mathcal{L}})=0$.
Finally, for $\dnplabel{$s$}$ we have
\begin{eqnarray*}
\fl
\bigl(\dnplabel{$s$},\boldsymbol{\mathcal{L}}\bigr]=
\{\dnnodolabel{$s$},\dnhlabel{$s$}{},\dnhlabel{}{$s$},\dnhllabel{$s$}{},
\dnhllabel{}{$s$},\dnhrlabel{$s$}{},\dnhrlabel{}{$s$},\dndoblelabel{$s$}{},
\dndoblelabel{}{$s$},\\
\dnclabel{$s$}{}{},\dnclabel{}{$s$}{},\dnclabel{}{}{$s$},
\dnllabel{$s$}{}{},\dnllabel{}{$s$}{},\dnllabel{}{}{$s$},\dnrlabel{$s$}{}{},
\dnrlabel{}{$s$}{},\dnrlabel{}{}{$s$},\boldsymbol{\mathcal{L}}\}
\end{eqnarray*}
which leads to $\mu_\mathcal{W}(\dnplabel{$s$},\boldsymbol{\mathcal{L}})=0$.
Substituting these values of the M\"obius function in
the general expression \eref{ec:generalfunctional3}, we recover
functional \eref{ec:finalfunctional}.

Although in our example we have shown that recursion~\ref{rec:rec1}
is enough to compute the M\"obius function, there are many alternative
(more efficient) techniques which exploit the order structure of the
cluster poset associated to the given lattice model and make use
of specialized results of the theory of posets (see section~3.8
of (Stanley~1999) and section~IV.3 of
(Aigner~1979)). In \ref{sec:app} we provide a few
examples of these tools.

\section{Extending lattice fundamental measure theory}
\label{sec:sec4b}
In the previous section we have achieved a reformulation of
LFMT based on theorem~\ref{theo:theo2}. Actually, we have shown that 
LFMT is a particular case of the approximation proposed in that 
theorem, where we choose the cluster set $\mathcal{W}_\mathrm{max}$
as the set of maximal \mbox{1-particle} cavities. We will see in the next 
section that, for hard-core models, this choice is an excellent 
balance between accuracy and simplicity of the approximate functional.
But we want to stress that the cluster expansion of the free-energy 
functional \eref{ec:clusterexpansion} is more general: it applies not
only to hard-core models, but is valid for \textsl{any} lattice model.

In order to illustrate this, let us consider the Ising lattice gas in 
an arbitrary lattice $\mathcal{L}$. The interaction potential for
this system is such that each lattice node can be occupied at most by 
one particle and two particles interact with an energy $J$ if they 
are placed at nodes which are nearest neighbors. If we applied 
theorem~\ref{theo:theo2} to this system with $\mathcal{W}_\mathrm{max}$
the set of maximal \mbox{1-particle} cavities (which for this model is just
the set of all single lattice nodes), the approximate free-energy functional
we would obtain would be that of the site-excluding ideal lattice gas,
i.e.
\begin{equation*}
\mathcal{F}_\mathrm{app}[\rho]=\sum_{\bi{s}\in\mathcal{L}}
\bigl[\rho(\bi{s})\ln\rho(\bi{s})+(1-\rho(\bi{s}))
\ln(1-\rho(\bi{s}))\bigr].
\end{equation*}
This choice for $\mathcal{W}_\mathrm{max}$ is clearly inappropriate 
for this system, since it ignores the interaction between nearest 
neighbors. In order to account for it we should go beyond the
standard LFMT and take 2-particle cavities. Thus,
$\mathcal{W}_\mathrm{max}=\{\mbox{all pairs of nearest
neighbors}\}$ and $\mathcal{W}=\{\mathcal{L}\}\cup
\{\mbox{all pairs of nearest neighbors}\}\cup\{\mbox{all single
nodes}\}$. The M\"obius function takes the values
$\mu_\mathcal{W}(\{\bi{s},\bi{t}\},\mathcal{L})=-1$
and $\mu_\mathcal{W}(\{\bi{s}\},\mathcal{L})=q(\bi{s})-1$
for every pair of nearest neighbors $\{\bi{s},\bi{t}\}$
and every single node $\{\bi{s}\}$ of $\mathcal{L}$,
$q(\bi{s})$ being the coordination number at node $\bi{s}$.
From theorem~\ref{theo:theo2}, the approximate free-energy functional
will be
\begin{equation}
\label{ec:isingfunctional}
\mathcal{F}_\mathrm{app}[\rho]=
\sum_{\mbox{\scriptsize all n.n.\ {}}\{\bi{s},\bi{t}\}}
\mathcal{F}_{\{\bi{s},\bi{t}\}}[\rho]-\sum_{\bi{s}\in\mathcal{L}}
[q(\bi{s})-1]\mathcal{F}_{\{\bi{s}\}}[\rho],
\end{equation}
where
\begin{eqnarray}
\label{ec:pairfunctional}
\fl
\mathcal{F}_{\{\bi{s},\bi{t}\}}[\rho]=\rho(\bi{s})
\ln\bigl[\rho(\bi{s})-\rho^{(2)}(\bi{s},\bi{t})\bigr]
+\rho(\bi{t}) \ln\bigl[\rho(\bi{t})-\rho^{(2)}(\bi{s},\bi{t})
\bigr]\nonumber \\
+\bigl[1-\rho(\bi{s})-\rho(\bi{t})\bigr]
\ln\bigl[1-\rho(\bi{s})-\rho(\bi{t})+\rho^{(2)}(\bi{s},\bi{t})\bigr],
\end{eqnarray}
$\rho^{(2)}(\bi{s},\bi{t})$ being the joint probability of finding
two particles at nodes $\bi{s}$ and $\bi{t}$. For this model it is
not difficult to show that it can be eliminated in terms of 
$\rho(\bi{s})$ and $\rho(\bi{t})$ as
\begin{equation*}
\fl
\rho^{(2)}(\bi{s},\bi{t})=\frac{1+\zeta[\rho(\bi{s})+\rho(\bi{t})]-
\sqrt{\bigl\{1+\zeta[\rho(\bi{s})+\rho(\bi{t})]\bigr\}^2
-4\zeta(1+\zeta)\rho(\bi{s})\rho(\bi{t})}}{2\zeta},
\end{equation*}
with $\zeta=\exp(-J)-1$ ($J$ is the interaction energy between
nearest neighbors in $kT$ units).
Note that $\mathcal{F}_{\{\bi{s}\}}[\rho]$ can be
obtained from \eref{ec:pairfunctional} by setting $\rho(\bi{t})=0$
and is just $\Phi_0\bigl(\brho(\bi{s})\bigr)$.
This approximation is exact when the lattice $\mathcal{L}$
is a Bethe lattice\footnote{In particular, it is exact for the
one-dimensional lattice $\mathbb{Z}$.} (even with node-dependent
coordination numbers) and it is equivalent to the Bethe
approximation for any other lattice (Bowman and Levin~1982).

\section{Low-density limit of the approximate functional}
\label{sec:sec5}

A fundamental ingredient of many approximate density functional theories
is the available exact information about the low-density limit
(Evans 1992; see also Cuesta \mbox{\etal{}2002} for an analysis of this
limit in the construction of the \textsl{weighted density approximation}
and the FMT). In contrast, in our formulation of LFMT we have input
the exact functionals for the system restricted to a certain set of
clusters and used a combinatorial tool (M\"obius inversion) to construct
an ``optimal'' functional out of them. The latter is so restrictive
that the only freedom we have in the resulting approximation
\eref{ec:finalappfunctional} is limited to the choice of 
$\mathcal{W}_\mathrm{max}$. In this section we will prove that with 
an appropriate choice of this set we can assure the correct behaviour
of the approximate functional in the low-density limit up to,
at least, third order. 

The second functional derivative of the excess part of the exact 
free-energy functional yields the exact pair direct correlation functional
\begin{eqnarray}
\fl
c^{(2)}_{ij}(\bi{s}_1,\bi{s}_2)&=
-\frac{\delta^2\mathcal{F}^{\mathrm{ex}}_{\boldsymbol{\mathcal{L}}}[\brho]}
{\delta\rho_i(\bi{s}_1)\delta\rho_j(\bi{s}_2)} \nonumber \\
\label{ec:correlationexpansion}
\fl
&\sim
f_{ij}(\bi{s}_1,\bi{s}_2)\left[1+\sum_{k}\sum_{\bi{s}_3\in\mathcal{L}}
f_{ik}(\bi{s}_1,\bi{s}_3)\rho_k(\bi{s}_3)f_{kj}(\bi{s}_3,\bi{s}_2)\right]
\quad (\brho\to \boldsymbol{0}),
\end{eqnarray}
where $f_{ij}(\bi{s}_1,\bi{s}_2)\equiv\rme^{-\phi_{ij}(\bi{s}_1,\bi{s}_2)}-1$
is the Mayer function, $\phi_{ij}(\bi{s}_1,\bi{s}_2)$ being the interaction 
potential (in $kT$ units) between a particle of species $i$ at node 
$\bi{s}_1$ and another
of species $j$ at node $\bi{s}_2$ (the interaction potential is assumed
pairwise). If we compute the pair direct correlation functional
from the approximate functional \eref{ec:finalappfunctional}, we obtain
\begin{equation*}
c^{(2)}_{\mathrm{app},ij}(\bi{s}_1,\bi{s}_2)=
\sum_{\boldsymbol{\mathcal{C}}\in\mathcal{W}-\{\boldsymbol{\mathcal{L}}\}}
\bigl[-\mu_\mathcal{W}(\boldsymbol{\mathcal{C}},\boldsymbol{\mathcal{L}})\bigr]
c^{(2)}_{ij}(\bi{s}_1,\bi{s}_2|\boldsymbol{\mathcal{C}}),
\end{equation*}
where we have introduced
\begin{equation*}
c^{(2)}_{ij}(\bi{s}_1,\bi{s}_2|\boldsymbol{\mathcal{C}})\equiv
-\frac{\delta^2\mathcal{F}^{\mathrm{ex}}_{\boldsymbol{\mathcal{C}}}[\brho]}
{\delta\rho_i(\bi{s}_1)\delta\rho_j(\bi{s}_2)},
\end{equation*}
the \textsl{exact} pair direct correlation functional for the system
restricted to cluster $\boldsymbol{\mathcal{C}}$. Since
$c^{(2)}_{ij}(\bi{s}_1,\bi{s}_2|\boldsymbol{\mathcal{C}})$ is exact, we
have from \eref{ec:correlationexpansion} that 
\begin{equation*}
c^{(2)}_{ij}(\bi{s}_1,\bi{s}_2|\boldsymbol{\mathcal{C}})
\sim
f_{ij}(\bi{s}_1,\bi{s}_2)\chi_{\mathcal{C}_i}(\bi{s}_1)
\chi_{\mathcal{C}_j}(\bi{s}_2)
\qquad(\brho\to\boldsymbol{0}),
\end{equation*}
with $\chi_{\mathcal{C}}(\bi{s})$ the indicator function of
$\mathcal{C}$. Thus, the approximate pair direct correlation
functional satisfies
\begin{equation*}
\fl
c^{(2)}_{\mathrm{app},ij}(\bi{s}_1,\bi{s}_2)\sim f_{ij}(\bi{s}_1,\bi{s}_2)
\sum_{\boldsymbol{\mathcal{C}}\in\mathcal{W}-\{\boldsymbol{\mathcal{L}}\}}
\bigl[-\mu_\mathcal{W}(\boldsymbol{\mathcal{C}},\boldsymbol{\mathcal{L}})\bigr]
\chi_{\mathcal{C}_i}(\bi{s}_1)\chi_{\mathcal{C}_j}(\bi{s}_2)
\qquad (\brho\to\boldsymbol{0}).
\end{equation*}
Taking into account that $\mu_\mathcal{W}(\boldsymbol{\mathcal{L}},
\boldsymbol{\mathcal{L}})=1$ and denoting $\bsigma_{ij}(\bi{s}_1,\bi{s}_2)
\equiv\bsigma_i(\bi{s}_1)\cup\bsigma_j(\bi{s}_2)$, we can
rewrite the above expression in the more suitable form
\begin{equation*}
c^{(2)}_{\mathrm{app},ij}(\bi{s}_1,\bi{s}_2)\sim f_{ij}(\bi{s}_1,\bi{s}_2)
\Bigl(1-\sum_{\bsigma_{ij}(\bi{s}_1,\bi{s}_2)\leq
\underset{\bbdot}{\boldsymbol{\mathcal{C}}}}
\mu_\mathcal{W}(\boldsymbol{\mathcal{C}},\boldsymbol{\mathcal{L}})\Bigr)
\qquad (\brho\to\boldsymbol{0}).
\end{equation*}
Now, inasmuch as $\bsigma_{ij}(\bi{s}_1,\bi{s}_2)\leq
\boldsymbol{\mathcal{C}}$ is equivalent to $\boldsymbol{\mathcal{C}}\cap
\bsigma_{ij}(\bi{s}_1,\bi{s}_2)=\bsigma_{ij}(\bi{s}_1,\bi{s}_2)$, it is a
direct consequence of corollary~\ref{cor:cor1} in~\ref{sec:app} that,
when $\brho\to\boldsymbol{0}$,
\begin{equation*}
\fl
c^{(2)}_{\mathrm{app},ij}(\bi{s}_1,\bi{s}_2)
\sim\cases{f_{ij}(\bi{s}_1,\bi{s}_2)&if 
$\bsigma_{ij}(\bi{s}_1,\bi{s}_2)$ is contained
in any cluster of $\mathcal{W}_\mathrm{max}$,\\
0&otherwise.}
\end{equation*}
The Mayer function has the same
range of the interaction potential, when the latter is short ranged,
$f_{ij}(\bi{s}_1,\bi{s}_2)$ vanishes if a particle of species
$i$ at $\bi{s}_1$ does not interact with a particle of species $j$
at $\bi{s}_2$. Therefore, if we want the approximate pair
direct correlation functional to recover the exact
low-density limit, it must happen that \textsl{for any pair of nodes
$\bi{s}_1$ and $\bi{s}_2$ for which $f_{ij}(\bi{s}_1,\bi{s}_2)\ne 0$
the cluster $\bsigma_{ij}(\bi{s}_1,\bi{s}_2)$ is
contained in at least one cluster of $\mathcal{W}_\mathrm{max}$}.

Let us consider the particular case of hard-core interaction.
For these systems, two particles interact if and only if they overlap.
Accordingly, we can define a 0D cavity as \textsl{any cluster such that
if we place two particles in any pair of its nodes} (of any allowed
species)\textsl{, they necessarily interact} (in which case the
corresponding Mayer function is non-zero). From this definition it
should be clear that if we let
$\mathcal{W}_\mathrm{max}$ be the set of maximal 0D cavities,
the density expansion of the approximate pair direct correlation
functional will recover the exact zeroth order.

Also, let us reconsider the Ising lattice gas. We showed in
\sref{sec:sec4b} that the choice of $\mathcal{W}_\mathrm{max}$ as the
set of maximal \mbox{1-particle} cavities gives rise to a very poor approximation
of the free-energy functional. Then, we took $\mathcal{W}_\mathrm{max}$
as the set of all pairs of nearest neighbors in order to account for
the interaction. This set contains all maximal 0D cavities according
to the new definition, and so we are certain that the free-energy
functional \eref{ec:isingfunctional} gives a pair direct correlation
functional with the exact zeroth order term
in the density expansion.
The new definition of 0D cavity that we have just introduced is thus
suitable for any kind of interaction (whether hard or soft, repulsive
or attractive) in the sense that choosing $\mathcal{W}_\mathrm{max}$
as the set of all maximal 0D cavities guarantees the correct low-density
limit of $c^{(2)}_{\mathrm{app},ij}(\bi{s}_1,\bi{s}_2)$.

Now, let us look at higher order terms in the density expansion.
If we take the third functional derivative of the excess
part of the free-energy functional, we obtain the
so-called triplet direct correlation functional
\begin{equation*}
\fl
c^{(3)}_{ijk}(\bi{s}_1,\bi{s}_2,\bi{s}_3)=-\frac{\delta^3
\mathcal{F}^{\mathrm{ex}}_{\boldsymbol{\mathcal{L}}}[\brho]}
{\delta\rho_i(\bi{s}_1)\delta\rho_j(\bi{s}_2)\delta\rho_k(\bi{s}_3)}
\sim f_{ij}(\bi{s}_1,\bi{s}_2) f_{jk}(\bi{s}_2,\bi{s}_3) f_{ki}(\bi{s}_3,\bi{s}_1)
\ \  (\brho\to\boldsymbol{0}).
\end{equation*}
Computing $c^{(3)}_{\mathrm{app},ijk}(\bi{s}_1,\bi{s}_2,\bi{s}_3)$
from \eref{ec:finalappfunctional} yields (reproducing the arguments
given to obtain $c^{(2)}_{\mathrm{app},ij}$), when
$\brho\to\boldsymbol{0}$,
\begin{equation*}
\fl
c^{(3)}_{ijk}(\bi{s}_1,\bi{s}_2,\bi{s}_3)
\sim f_{ij}(\bi{s}_1,\bi{s}_2) f_{jk}(\bi{s}_2,\bi{s}_3)
f_{ki}(\bi{s}_3,\bi{s}_1)
\biggl(1-\sum_{\bsigma_{ijk}(\bi{s}_1,\bi{s}_2,\bi{s}_3)\leq
\underset{\bbdot}{\boldsymbol{\mathcal{C}}}}
\mu_\mathcal{W}(\boldsymbol{\mathcal{C}},\boldsymbol{\mathcal{L}})\biggr),
\end{equation*}
where $\bsigma_{ijk}(\bi{s}_1,\bi{s}_2,\bi{s}_3)\equiv\bsigma_i(\bi{s}_1)\cup
\bsigma_j(\bi{s}_2)\cup\bsigma_k(\bi{s}_3)$. Therefore, we are guaranteed to
recover the exact low-density limit of the triplet direct correlation functional
if \textsl{every cluster $\bsigma_{ijk}(\bi{s}_1,\bi{s}_2,\bi{s}_3)$
for which the product $f_{ij}(\bi{s}_1,\bi{s}_2) f_{jk}(\bi{s}_2,\bi{s}_3)
f_{ki}(\bi{s}_3,\bi{s}_1)$ does not vanish is contained in at least one
cluster of $\mathcal{W}_\mathrm{max}$.} Note that because of
\eref{ec:correlationexpansion}, if this holds then the density expansion of 
the pair direct correlation functional is exact up to first order in
$\brho$.

Again, the choice of $\mathcal{W}_\mathrm{max}$ as the set of maximal 0D 
cavities (according to the new definition) is enough to assure the correct 
behaviour of $c^{(3)}_{\mathrm{app},ijk}(\bi{s}_1,\bi{s}_2,\bi{s}_3)$
in the low-density limit. To see this, just note that the product 
$f_{ij}(\bi{s}_1,\bi{s}_2) f_{jk}(\bi{s}_2,\bi{s}_3) f_{ki}(\bi{s}_3,\bi{s}_1)$
is different from zero only if two particles of the corresponding species 
placed at any pair of nodes $\{\bi{s}_1,\bi{s}_2,\bi{s}_3\}$ interact; in
other words, only if the nodes belong to the same maximal 0D cavity.
Therefore, \textsl{the approximate free-energy functional
\eref{ec:finalappfunctional} with $\mathcal{W}_\mathrm{max}$ the set of
maximal 0D cavities recovers the exact density expansion of the pair and
triplet direct correlation functional up to first and zeroth order, respectively}.

All this analysis can in principle be extended to higher-order direct
correlation functions. In general, we will have that
$c^{(n)}_{\mathrm{app},i_1\ldots i_n}(\bi{s}_1,\ldots,\bi{s}_n)$
will recover the exact low-density limit provided
the cluster $\bsigma_{i_1\ldots i_n}(\bi{s}_1,\ldots,\bi{s}_n)
\equiv\bigcup_{l=1}^n\bsigma_{i_l}(\bi{s}_l)$
is contained in at least one cluster of $\mathcal{W}_\mathrm{max}$
for any combination of $\{(i_1,\bi{s}_1),\ldots,(i_n,\bi{s}_n)\}$
for which the exact low-density limit is different from zero.
In practice, this is a very demanding task, because guaranteeing
the correct low-density limit of
$c^{(n)}_{\mathrm{app},i_1\ldots i_n}(\bi{s}_1,\ldots,\bi{s}_n)$
amounts to consider all diagrams in the virial expansion of 
the $n$th order direct correlation function, and the number of
them grows exponentially. Clearly, the current definition of 0D 
cavity is insufficient to provide the correct behaviour beyond $n=3$
(except in very particular cases, like some one-dimensional systems
for which the approximation becomes exact), so if we want more terms
in the low-density expansion we have to take bigger maximal
0D cavities. Bigger cavities means more particles in one cavity
and an increasingly higher difficulty to obtain the exact free-energy
functional for a single cavity. So, as usual with expansions, although
a systematic improvement is possible, going beyond the lowest
terms may be too involved in practice.

\section{Dimensional reduction}
\label{sec:sec5a}
\begin{figure}
\begin{center}
\includegraphics*[width=70mm]{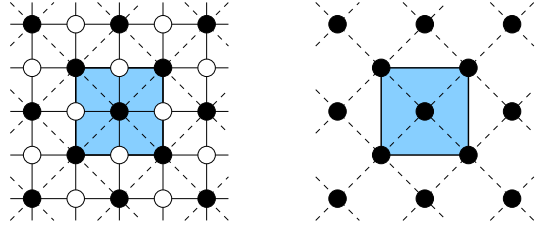}
\caption[]{\label{fig:fig2} The system on the left corresponds to
the lattice gas with first- and second-neighbor exclusion in
the square lattice. If the position of the particles is confined
to the black nodes of the lattice, the system behaves exactly like
the lattice gas with nearest-neighbor exclusion in
the square lattice (represented in the right figure).}
\end{center}
\end{figure}

One of the most remarkable properties of FMT is dimensional crossover.
This property means that by confining the particles of a
$d$-dimensional system to lie in a $(d-1)$-dimensional subset we
obtain the $(d-1)$-dimensional FM functional out of the
$d$-dimensional one. In the case of LFMT, a typical example would be
to start with the lattice gas with nearest-neighbor exclusion in the
simple cubic lattice in three dimensions and constrain the position of
the particles to the nodes of one of the coordinate planes in order
to obtain an effective system equivalent to the lattice
gas with nearest-neighbor exclusion in the square lattice.
In this example, the real dimension of the system is reduced
from three to two, and in so doing, the FM functional of the
three-dimensional system is transformed into that of the
two-dimensional one (Lafuente and Cuesta 2003). But this is only
an instance of a more general class of mappings between different
models. As another example, the FM of the lattice gas with nearest-neighbor
exclusion in the square lattice can also be obtained from the one of the 
lattice gas with first- and second-neighbor exclusion in the same lattice,
as \fref{fig:fig2} illustrates (Lafuente and Cuesta 2003).

The aim of this section is to define a general mapping between models
(of which these two examples are particular cases) and to prove that
the approximation \eref{ec:finalappfunctional} is ``closed'' with
respect to this kind of mapping, i.e.\ that the FM functional
of the original model becomes the one of the transformed model under
the action of that mapping on the density profile.

Let us assume that the original system has an underlying
lattice $\mathcal{L}$ and its approximate free-energy functional
is given from theorem~\ref{theo:theo2} by
\begin{equation}
\label{ec:originalfunctional}
\mathcal{F}_\mathrm{app}[\brho]=\sum_{\boldsymbol{\mathcal{C}}\in
\mathcal{W}-\{\boldsymbol{\mathcal{L}}\}}\bigl[-\mu_\mathcal{W}
(\boldsymbol{\mathcal{C}},\boldsymbol{\mathcal{L}})\bigr]
\mathcal{F}_{\boldsymbol{\mathcal{C}}}[\brho]
\end{equation}
for a given choice of $\mathcal{W}_\mathrm{max}$.
If we restrict the position of the particles of species $i$ to the
embedded lattice $\mathcal{L}'_i\subset\mathcal{L}$ (lattice here is
a general term which refers to any subset of $\mathcal{L}$, finite
or infinite), the approximate functional of the new effective system
can be obtained by specializing functional \eref{ec:originalfunctional}
to a density profile with support $\boldsymbol{\mathcal{L}'}=
\bigl(\mathcal{L}'_1,\ldots,\mathcal{L}'_p\bigr)$, i.e.
\begin{equation*}
\mathcal{F}'_\mathrm{app}[\brho]=
\mathcal{F}_\mathrm{app}[\brho_{\boldsymbol{\mathcal{L}'}}]=
\sum_{\boldsymbol{\mathcal{C}}\in\mathcal{W}-\{\boldsymbol{\mathcal{L}}\}}
\bigl[-\mu_\mathcal{W}(\boldsymbol{\mathcal{C}},\boldsymbol{\mathcal{L}})
\bigr] \mathcal{F}_{\boldsymbol{\mathcal{C}}\cap\boldsymbol{\mathcal{L}'}}
[\brho_{\boldsymbol{\mathcal{L}'}}],
\end{equation*}
where $\brho_{\boldsymbol{\mathcal{L}'}}(\bi{s})=\brho(\bi{s})$ if
$\bi{s}\in\boldsymbol{\mathcal{L}'}$ and is $0$ otherwise.
Since $\mathcal{F}_\varnothing[\brho]=0$, if $\mathcal{W}'$
denotes the set of all non-empty intersections of the
clusters in $\mathcal{W}$
with the cluster $\boldsymbol{\mathcal{L}'}$,
the above expression becomes
\begin{equation}
\label{ec:dr1}
\mathcal{F}'_\mathrm{app}[\brho]=\sum_{\boldsymbol{\mathcal{C}'}\in
\mathcal{W}'}\mathcal{F}_{\boldsymbol{\mathcal{C}'}}[\brho]
\sum_{\underset{\bbdot}{
\boldsymbol{\mathcal{C}}}\cap\boldsymbol{\mathcal{L}'}=
\boldsymbol{\mathcal{C}'}}[-\mu_\mathcal{W}(\boldsymbol{\mathcal{C}},
\boldsymbol{\mathcal{L}})+\delta(\boldsymbol{\mathcal{C}},
\boldsymbol{\mathcal{L}})\bigr],
\end{equation}
where we have introduced the identity of $I(\mathcal{W},
\mathbb{K})$, $\delta(\boldsymbol{\mathcal{C}},
\boldsymbol{\mathcal{L}})$, to compensate for the inclusion of 
$\mu_\mathcal{W}(\boldsymbol{\mathcal{L}},\boldsymbol{\mathcal{L}})=1$
in the sum.
We can now apply corollary~\ref{cor:cor1} in~\ref{sec:app}
to evaluate the constrained sum on the r.h.s.\ of \eref{ec:dr1}.
If $\boldsymbol{\mathcal{L}'}$ is contained in some cluster of
$\mathcal{W}-\{\boldsymbol{\mathcal{L}}\}$, then we recover the result
of section~\ref{sec:sec3}, i.e.\ $\mathcal{F}'_\mathrm{app}[\brho]=
\mathcal{F}_{\boldsymbol{\mathcal{L}'}}[\brho]$ (in other words, the
approximate functional of the effective system coincides with the exact
one). On the contrary, if $\boldsymbol{\mathcal{L}'}$ is not contained in
any cluster of $\mathcal{W}-\{\boldsymbol{\mathcal{L}}\}$, then
\begin{equation}
\label{ec:dr2}
\mathcal{F}'_\mathrm{app}[\brho]=
\sum_{\boldsymbol{\mathcal{C}'}\in\mathcal{W}'-\{\boldsymbol{\mathcal{L}'}\}}
\bigl[-\mu_{\mathcal{W}'}(\boldsymbol{\mathcal{C}'},\boldsymbol{\mathcal{L}'})\bigr]
\mathcal{F}_{\boldsymbol{\mathcal{C}'}}[\brho].
\end{equation}
Note that if $\mathcal{W}_\mathrm{max}$ denotes the set of maximal
clusters of $\mathcal{W}-\{\boldsymbol{\mathcal{L}}\}$, then
the set of maximal clusters of $\mathcal{W}'-\{\boldsymbol{\mathcal{L}'}\}$
is the set $\mathcal{W}'_\mathrm{max}$ of all non-empty intersections of
the clusters of $\mathcal{W}_\mathrm{max}$ with $\boldsymbol{\mathcal{L}'}$.
Consequently, the set $\mathcal{W}'$ coincides with the set of all non-empty
intersections of the clusters in $\mathcal{W}'_\mathrm{max}$,
as well as the cluster $\boldsymbol{\mathcal{L}'}$. In other words:
the approximate functional \eref{ec:dr2} is the one
that would have been obtained directly from theorem~\ref{theo:theo2}
if it had been applied to the effective system (the original system
constrained to $\boldsymbol{\mathcal{L}'}$) choosing $\mathcal{W}'_\mathrm{max}$
as the set of maximal clusters.

The previous result shows that LFMT is closed under dimensional
reduction. Note that if we have the approximate free-energy functional
prescribed by theorem~2 for a given system and a certain
$\mathcal{W}_\mathrm{max}$, then for all systems that can be
obtained from it through dimensional reduction, the resulting
approximate functional for the lower-dimensional system coincides 
with the one deduced from theorem~2 with $\mathcal{W}'_\mathrm{max}$
obtained from $\mathcal{W}_\mathrm{max}$ as above. In LFMT, the
prescription for $\mathcal{W}_\mathrm{max}$ for a given system is
to choose all maximal 0D cavities. In order for LFMT to be closed
under dimensional reduction, $\mathcal{W}'_\mathrm{max}$ should
coincide with the set of maximal 0D cavities of the lower-dimensional
system. A little reflection will convince the reader that this
is indeed the case (and it is so as well for other kind of mappings,
like the one described in \fref{fig:fig2}). In general, if a system
$S$ is transformed by the action of a mapping $\mathcal{T}$ into
another system $\mathcal{T}\{S\}$, and if its set of maximal clusters
$\mathcal{W}_\mathrm{max}(S)$ is transformed as described above
into the set $\mathcal{W}'_\mathrm{max}=\mathcal{T}\big\{
\mathcal{W}_\mathrm{max}(S)\big\}$, then the approximate functional
\eref{ec:finalappfunctional} behaves consistently under
$\mathcal{T}$ provided the prescription to choose 
$\mathcal{W}_\mathrm{max}(S)$ is such that
$\mathcal{W}_\mathrm{max}\big(\mathcal{T}\{S\}\big)=
\mathcal{T}\big\{\mathcal{W}_\mathrm{max}(S)\big\}$.

\section{Conclusions}

In this work we have provided a rigorous foundation of LFMT
based on the formalism of M\"obius inversion in posets of
lattice clusters. The free-energy density functional is thus
expressed in the form of a cluster expansion: given a set
of basic clusters $\mathcal{W}_\mathrm{max}$, there is a
unique functional of the form \eref{ec:finalappfunctional}
which is exact for every density profile with support any
subcluster of $\mathcal{W}_\mathrm{max}$. The cluster
expansion \eref{ec:finalappfunctional} requires the
exact expressions of the free-energy functional on the
clusters of $\mathcal{W}_\mathrm{max}$. The low-density
limit of the functional \eref{ec:finalappfunctional} dictates
a definition of the clusters of $\mathcal{W}_\mathrm{max}$ 
---the 0D cavities, or those clusters such that if there are
two or more particles in them they necessarily interact---
which guarantees the exact zero-density limit of the pair
and triplet direct correlation functions. This redefinition
subsumes the previous version of LFMT, valid for hard-core
models (Lafuente and Cuesta~2004), and extends it to include
any lattice model with short-range interaction (the Ising
lattice gas is an explicit example). The M\"obius function
formalism also allows us to analyze the behaviour of
functional \eref{ec:finalappfunctional} under mappings
between models, and a consequence of this analysis is the
proof that LFMT behaves consistently under dimensional
reduction (or confinements of the density into lower-dimensional
sets of the lattice).

Going beyond LFMT ---to account for higher order correlations,
for instance--- is, in principle, possible, but requires
choices of $\mathcal{W}_\mathrm{max}$ containing larger
clusters. But then the exact free-energy functional of these
larger clusters is required, what may be too involved. In
some cases, though ---\textsl{e.g.} in some one-dimensional
models---, it is known that LFMT as such is exact, so no
improvement is required in those cases. The reason behind
this fact is not clear to us yet, and it is certainly a
matter that deserves further thought.

Another interesting point concerns the continuum limit.
For some models \textsl{(e.g.} hard cubes in a
$d$-dimensional cubic lattice) this limit is feasible and
in fact the limit functional coincides with the FM functional
of the continuum model (Lafuente~2004).
But there are important cases, like
hard spheres, which are not easy to obtain as a limit of
discrete lattice models. If this were possible, the result
would be a functional which would recover the exact zero-density
limit of the pair and triplet direct correlation functions,
something that the best current FM functional for hard
spheres does not accomplish (Tarazona and Rosenfeld~1997,
Tarazona~2000, Cuesta \etal{}~2002). From the insight provided
by the present analysis of LFMT we dare to say that 
Tarazona's functional for hard spheres (Tarazona~2000) is
the limit of the functionals \eref{ec:finalappfunctional}
of a sequence of lattice models with an incomplete choice
of $\mathcal{W}_\mathrm{max}$ (incomplete in the sense that
some maximal 0D cavities are not in $\mathcal{W}_\mathrm{max}$).
This is another problem certainly worth exploring.

\ack
This work is supported by project BFM2003--0180 of the
Spanish Ministerio de Ciencia y Tecnolog\'{\i}a.

\appendix

\section{Some technical results}
\label{sec:app}
In order to make this work self-contained, in this appendix we
will collect and prove some technical results of the theory of
posets (and in particular of cluster posets) that we have used
along this article. We will only prove those which are original
contributions, while for the rest we simply address the reader
to the specialized literature, where complete proofs can be found.

In \sref{sec:sec3} we have proved that the approximate
functional \eref{ec:finalappfunctional} is exact when it
is restricted to any cluster of the set $\mathcal{W}-
\{\boldsymbol{\mathcal{L}}\}$. Also, in sections~\ref{sec:sec5}
and \ref{sec:sec5a}
we have analyzed the low-density limit and the dimensional crossover
of the functional \eref{ec:finalappfunctional},
respectively. In all cases, the
key point was to evaluate a constrained sum of the M\"obius 
function $\mu_\mathcal{W}(\boldsymbol{\mathcal{C}},
\boldsymbol{\mathcal{L}})$ for those $\boldsymbol{\mathcal{C}}\in
\mathcal{W}$ satisfying $\boldsymbol{\mathcal{C}}\cap
\boldsymbol{\mathcal{D}}=\boldsymbol{\mathcal{E}}$,
$\boldsymbol{\mathcal{D}}$ and $\boldsymbol{\mathcal{E}}$ being
clusters of the lattice $\boldsymbol{\mathcal{L}}$ not necessarily
in $\mathcal{W}$. This relation can be expressed in the more
general form $\sigma^{+}\boldsymbol{\mathcal{C}}=
\boldsymbol{\mathcal{E}}$, where $\sigma^{+}$ is a mapping from
the cluster poset $\mathcal{W}$ to another cluster poset which,
in general, is different from $\mathcal{W}$. For all instances
in this work, $\sigma^{+}$ is just the intersection with a
given fixed cluster of the underlying lattice, and hence
is an order-preserving map.

The advantage of the above interpretation of the constraint over the
clusters in $\mathcal{W}$ is that it allows to use an important
result of the theory of posets which relates the M\"obius functions
of two posets, $\mathcal{W}$ and $\mathcal{V}$, if both are connected
through a pair $(\sigma,\sigma^{+})$ of order-preserving maps,
where $\sigma$ is a \textsl{Galois function}, defined as

\begin{defi}
\label{defi:defi1}
Let $\mathcal{V}$ and $\mathcal{W}$ be posets. A mapping
$\sigma\colon \mathcal{V}\rightarrow \mathcal{W}$ is called a
\textsl{Galois function} if there exists a function $\sigma^{+}\colon
\mathcal{W} \rightarrow \mathcal{V}$ such that
\begin{enumerate}
\item $\sigma,\, \sigma^{+}$ are order-preserving maps;
\item $\sigma^{+}\sigma x \geq x$ for all $x\in \mathcal{V}$ and
$\sigma\sigma^{+} z \leq z$ for all $z\in \mathcal{W}$.
\end{enumerate}
\end{defi}

Given two posets $\mathcal{V}$ and $\mathcal{W}$, their M\"obius
functions are related via a Galois function in the precise way
expressed in the following theorem (which is a version
of Theorem~4.39 in p.~173 of Aigner (1979)):

\begin{theo}
\label{theo:theo3}
Let $\sigma\colon \mathcal{V}\rightarrow \mathcal{W}$ be a Galois
function. Then, for all $x\in \mathcal{V}$, $y\in \mathcal{W}$,
\begin{equation*}
\fl
\sum_{z\in \mathcal{W},\sigma^{+}z=x}\mu_{\mathcal{W}}(z,y)=\cases{
\mu_{\sigma^{+}\mathcal{W}}(x,\sigma^{+}y)&if $x\in\sigma^{+} \mathcal{W}$
and $y\in\sigma \mathcal{V}$,\\
0& otherwise.}
\end{equation*}
\end{theo}

This theorem is exactly what we need to derive a result which
can be applied directly to evaluate the constrained sums that
appear along this article:

\begin{coro}
\label{cor:cor1}
Let $\mathcal{W}$ be a cluster poset with underlying lattice $\mathcal{L}$
which contains all non-empty intersections of certain cluster poset
$\mathcal{W}_\mathrm{max}$ as well as the cluster $\boldsymbol{\mathcal{L}}$.
Let $\boldsymbol{\mathcal{L}'}$ be a cluster of lattice $\mathcal{L}$ not
necessarily in $\mathcal{W}$, and let us consider the poset
$\mathcal{W}'$ of all non-empty intersections of the clusters
in $\mathcal{W}$ with $\boldsymbol{\mathcal{L}'}$. Then, for all
$\boldsymbol{\mathcal{C}'}\in\mathcal{W}'$,
\begin{equation*}
\fl
\sum_{\boldsymbol{\mathcal{C}}\in\mathcal{W},\,\boldsymbol{\mathcal{C}}\cap
\boldsymbol{\mathcal{L}'}=\boldsymbol{\mathcal{C}'}}
\mu_\mathcal{W}(\boldsymbol{\mathcal{C}},
\boldsymbol{\mathcal{L}})=\cases{\mu_{\mathcal{W}'}(\boldsymbol{\mathcal{C}'},
\boldsymbol{\mathcal{L}'})&if $\boldsymbol{\mathcal{L}'}\not\subset
\boldsymbol{\mathcal{D}}$ for all $\boldsymbol{\mathcal{D}}\in
\mathcal{W}-\{\boldsymbol{\mathcal{L}}\}$,\\
0& otherwise.}
\end{equation*}
\end{coro}

\textsf{\textsl{Proof.}} 
To apply theorem~3, we first need to rewrite the constrained sum in
the statement of the corollary in a more convenient form. Since
$\boldsymbol{\varnothing}\notin\mathcal{W}'$, the only clusters
in $\mathcal{W}$ contributing to the sum are those whose intersection
with $\boldsymbol{\mathcal{L}'}$ is non-empty. Let
$\widetilde{\mathcal{W}}$ denote the poset of such clusters; then
we have that $\mu_{\mathcal{W}}(\boldsymbol{\mathcal{C}},\boldsymbol{\mathcal{L}})=
\mu_{\widetilde{\mathcal{W}}}(\boldsymbol{\mathcal{C}},\boldsymbol{\mathcal{L}})$
for all $\boldsymbol{\mathcal{C}}\in \widetilde{\mathcal{W}}$, since
$[\boldsymbol{\mathcal{C}},\boldsymbol{\mathcal{L}}]$ in $\mathcal{W}$
is identical to $[\boldsymbol{\mathcal{C}},\boldsymbol{\mathcal{L}}]$
in $\widetilde{\mathcal{W}}$. Therefore we have the identity
\begin{equation}
\label{ec:cor1}
\sum_{\boldsymbol{\mathcal{C}}\in\mathcal{W},\,
\boldsymbol{\mathcal{C}}\cap\boldsymbol{\mathcal{L}'}=\boldsymbol{\mathcal{C}'}}
\mu_\mathcal{W}(\boldsymbol{\mathcal{C}},\boldsymbol{\mathcal{L}})=
\sum_{\boldsymbol{\mathcal{C}}\in\widetilde{\mathcal{W}},\,
\sigma^{+}\boldsymbol{\mathcal{C}}=\boldsymbol{\mathcal{C}'}}
\mu_{\widetilde{\mathcal{W}}}(\boldsymbol{\mathcal{C}},\boldsymbol{\mathcal{L}}),
\end{equation}
where $\sigma^{+}\colon \widetilde{\mathcal{W}}\rightarrow \mathcal{W}'$
is defined as $\sigma^{+}\boldsymbol{\mathcal{C}}\equiv
\boldsymbol{\mathcal{C}}\cap\boldsymbol{\mathcal{L}'}$.

Now, in order to apply theorem~\ref{theo:theo3} to \eref{ec:cor1}
we need a pair $(\sigma,\sigma^{+})$ of order-preserving maps
$\sigma\colon \mathcal{W}'\rightarrow\widetilde{\mathcal{W}}$ and
$\sigma^+\colon\widetilde{\mathcal{W}}\rightarrow\mathcal{W}'$,
$\sigma$ being a Galois function. Let us take for $\sigma^+$ the
one introduced in \eref{ec:cor1} and let us define $\sigma$ as
$\sigma\boldsymbol{\mathcal{C}'}\equiv\inf\{\boldsymbol{\mathcal{C}}
\in\widetilde{\mathcal{W}}\,|\,\boldsymbol{\mathcal{C}'}
\subset\boldsymbol{\mathcal{C}}\}$
for all $\boldsymbol{\mathcal{C}'}\in\mathcal{W}'$. It is a direct
consequence of the definition that $\sigma$ and $\sigma^+$ both are
order-preserving maps. Moreover, for all $\boldsymbol{\mathcal{C}}\in
\widetilde{\mathcal{W}}$ we have
\begin{equation*}
\sigma \sigma^{+} \boldsymbol{\mathcal{C}}=\inf\{\boldsymbol{\mathcal{D}}\in
\widetilde{\mathcal{W}}\,|\, \sigma^+\boldsymbol{\mathcal{C}}\subset
\boldsymbol{\mathcal{D}}\}\leq\boldsymbol{\mathcal{C}}
\end{equation*}
because $\sigma^+\boldsymbol{\mathcal{C}}\subset\boldsymbol{\mathcal{C}}$,
and for all $\boldsymbol{\mathcal{C}'}\in \mathcal{W}'$, 
\begin{equation*}
\sigma^+ \sigma \boldsymbol{\mathcal{C}'}=
\sigma^+\inf\{\boldsymbol{\mathcal{D}}\in\widetilde{\mathcal{W}}\,|\,
\boldsymbol{\mathcal{C}'}\subset\boldsymbol{\mathcal{D}}\}=
\inf\{\boldsymbol{\mathcal{D}'}\in\mathcal{W}'\,|\,
\boldsymbol{\mathcal{C}'}\subset\boldsymbol{\mathcal{D}'}\}=
\boldsymbol{\mathcal{C}'}.
\end{equation*}
Therefore, $\sigma$ and $\sigma^+$ satisfy condition (ii) of
definition~\ref{defi:defi1} and thus $\sigma$ is a Galois function.  

At this point, we can apply theorem~\ref{theo:theo3} to the r.h.s of
\eref{ec:cor1}. Since for this case $\sigma^+\widetilde{\mathcal{W}}=\mathcal{W}'$,
for all $\boldsymbol{\mathcal{C}'}\in\mathcal{W}'$ we can write 
\begin{equation*}
\sum_{\boldsymbol{\mathcal{C}}\in\mathcal{W},\,\boldsymbol{\mathcal{C}}\cap
\boldsymbol{\mathcal{L}}=\boldsymbol{\mathcal{C}'}}
\mu_\mathcal{W}(\boldsymbol{\mathcal{C}},\boldsymbol{\mathcal{L}})=\cases{
\mu_{\mathcal{W}'}(\boldsymbol{\mathcal{C}'},\boldsymbol{\mathcal{L}'})&if
$\boldsymbol{\mathcal{L}}\in\sigma\mathcal{W}'$,\\
0&otherwise.}
\end{equation*}
The last step of the proof just amounts to showing that
$\boldsymbol{\mathcal{L}}\in\sigma\mathcal{W}'$ if and only
if $\boldsymbol{\mathcal{L}'}$ is not contained in
any cluster of $\mathcal{W}-\{\boldsymbol{\mathcal{L}}\}$.
Note that the latter is equivalent to
$\sigma\boldsymbol{\mathcal{L}'}=\boldsymbol{\mathcal{L}}$,
therefore if it holds then $\boldsymbol{\mathcal{L}}\in\sigma\mathcal{W}'$.
Now, let us assume that $\boldsymbol{\mathcal{L}}\inḉ\sigma
\mathcal{W}'$, then there exists a cluster $\boldsymbol{\mathcal{C}'}$
in $\mathcal{W}'$ such that $\sigma\boldsymbol{\mathcal{C}'}=
\boldsymbol{\mathcal{L}}$. Since $\sigma$ is order-preserving
and $\boldsymbol{\mathcal{C}'}\leq\boldsymbol{\mathcal{L}'}$,
we have $\boldsymbol{\mathcal{L}}=\sigma\boldsymbol{\mathcal{C}'}\leq
\sigma\boldsymbol{\mathcal{L}'}$, where only the equality
$\sigma\boldsymbol{\mathcal{L}'}=\boldsymbol{\mathcal{L}}$ can
hold, and the proof is complete. $\blacksquare$

To end this appendix we will bring about some results which simplify
the calculation of the M\"obius function of certain posets, and
we will provide two applications related to the cluster posets
involved in this work. First of all, we will give some definitions
relative to a special type of posets: \textsl{lattices}
(a mathematical notion not to be confused with physical lattices),
since they appear in a natural way when we have to compute the
M\"obius function of a locally finite poset.

A poset $\mathcal{P}$ is a \textsl{lattice} if for any $x,y\in\mathcal{P}$,
$\sup\{x,y\}$ and $\inf\{x,y\}$ are in $\mathcal{P}$. One
instance of a lattice is any interval $[\boldsymbol{\mathcal{C}},
\boldsymbol{\mathcal{C}'}]$ of a cluster poset $\mathcal{W}$
such as those involved in theorem~\ref{theo:theo2} (which are
closed under non-empty intersections). A 0-element of a poset
$\mathcal{W}$, denoted $\hat{0}$, is an element satisfying
$\hat{0}\leq x$ for all $x\in\mathcal{W}$. Dually,
a 1-element of $\mathcal{W}$, denoted $\hat{1}$, is
an element satisfying $x\leq \hat{1}$ for all $x\in\mathcal{W}$.
Obviously, any finite lattice has a $\hat{0}$ and a $\hat{1}$.
A \textsl{point} of a finite lattice $\mathcal{P}$ is an element
satisfying $\hat{0}<x$ for which there is no element $z\in\mathcal{P}$
such that $z<x$. A \textsl{copoint} of a finite lattice $\mathcal{P}$
is an element satisfying $x<\hat{1}$ for which there is no element
$z\in\mathcal{P}$ such that $x<z$. A subset $\mathcal{M}$
of a finite lattice $\mathcal{P}$ is called a \textsl{lower
cross-cut} if $\hat{0}\notin \mathcal{M}$ and for
all $\hat{0}\neq x\in\mathcal{P}$ with $x\notin\mathcal{M}$
there is an element $y\in\mathcal{M}$ with $y\leq x$.
Dually, a subset $\mathcal{M}$ is an \textsl{upper cross-cut}
if $\hat{1}\notin\mathcal{M}$ and for all $\hat{1}\neq x\in\mathcal{P}$
with $x\notin\mathcal{M}$ there is an element $y\in\mathcal{M}$
with $x\leq y$. An example of lower (upper) cross-cut is the
set of all points (copoints) of a given finite lattice.

In our particular case, we have to compute the M\"obius function
$\mu_\mathcal{W}(\boldsymbol{\mathcal{C}},\boldsymbol{\mathcal{L}})$,
where $\mathcal{W}$ is a cluster poset which contains
all non-empty intersections of the elements of certain
cluster set $\mathcal{W}_\mathrm{max}$ as well as
the cluster $\boldsymbol{\mathcal{L}}$, which is a 1-element.
Let us consider the finite lattice $\mathcal{P}_{\boldsymbol{\mathcal{C}}}
\equiv [\boldsymbol{\mathcal{C}},\boldsymbol{\mathcal{L}}]$ with
$\hat{0}=\boldsymbol{\mathcal{C}}$ and $\hat{1}=\boldsymbol{\mathcal{L}}$.
From recursion~\ref{rec:rec1}, it is straightforward that
for all $\boldsymbol{\mathcal{C}}\inḿ\mathcal{W}$ we have
$\mu_\mathcal{W}(\boldsymbol{\mathcal{C}},\boldsymbol{\mathcal{L}})=
\mu_{\mathcal{P}_{\boldsymbol{\mathcal{C}}}}(\hat{0},\hat{1})$.
Therefore, in all cases we have to compute the M\"obius
function $\mu_{\mathcal{P}}(\hat{0},\hat{1})$ of certain
finite lattice $\mathcal{P}$. This fact makes the following
theorem (from Aigner (1979), Theorem~4.42 on p.~175)
very useful:

\begin{theo}\textbf{\textsf{(Cross-cut theorem)}}
Let $\mathcal{P}$ be a finite lattice, $\mathcal{M}$ a lower
(upper) cross-cut and $n_k$ the number of sets
$\mathcal{A}\subset\mathcal{M}$ with $k$ elements such that
$\sup \mathcal{A}=\hat{1}$ ($\inf\mathcal{A}=\hat{0}$). Then
\begin{equation*}
\mu_{\mathcal{P}}(\hat{0},\hat{1})=\sum_{k\geq 0} (-1)^k n_k.
\end{equation*}
\end{theo} 

When we take in this theorem the lower (upper) cross-cut
as the set of all points (copoints) of the lattice $\mathcal{P}$,
a direct consequence is the following corollary (from Stanley (1999),
Corollary~3.9.5 on p.~126):

\begin{coro}
\label{cor:cor2}
Let $\mathcal{P}$ be a finite lattice with point set $\mathcal{Q}$
and copoint set $\mathcal{R}$. Then
\begin{equation*}
\mu_\mathcal{P}(\hat{0},\hat{1})=0
\end{equation*}
if $\hat{0}\neq \inf\mathcal{R}$ or $\hat{1}\neq \sup \mathcal{Q}$.
\end{coro}
Note that this result renders the calculation of some values of
the M\"obius function of a given poset straightforward.

A practical application of corollary~\ref{cor:cor2}
is to easily compute some values of the M\"obius function, say,
in the working example of section~\ref{sec:sec4}.
Note that it is a direct consequence of this corollary
that $\mu_\mathcal{W}(\dnhllabel{$s$}{},\boldsymbol{\mathcal{L}})=
\mu_\mathcal{W}(\dnhrlabel{$s$}{},\boldsymbol{\mathcal{L}})=
\mu_\mathcal{W}(\dnhlabel{$s$}{},\boldsymbol{\mathcal{L}})=
\mu_\mathcal{W}(\dnnodolabel{$s$}{},\boldsymbol{\mathcal{L}})=
\mu_\mathcal{W}(\dnplabel{$s$}{},\boldsymbol{\mathcal{L}})=0$
for any $s\in\mathcal{L}$, since for all these clusters the
supremum of the sets of points of the corresponding
interval is different from $\boldsymbol{\mathcal{L}}$
(a glance at \fref{fig:fig1} is enough to realize it).

An important consequence of corollary~\ref{cor:cor2} concerns the
form of the cluster expansion of the free-energy functional.
Notice that in our formulation of LFMT we have worked with
$\mathcal{W}$ defined as the cluster
$\boldsymbol{\mathcal{L}}$ and the non-empty intersections of the
clusters in $\mathcal{W}_\mathrm{max}$. We have shown that this
choice is enough to ensure our main purpose, building an approximate
free-energy functional which is exact in the clusters of
$\mathcal{W}_\mathrm{max}$. Having this idea in mind we could have
started with the cluster poset $\mathcal{V}$ made of cluster
$\boldsymbol{\mathcal{L}}$ and all non-empty clusters contained in
some cluster of $\mathcal{W}_\mathrm{max}$ (note that
$\mathcal{W}\subset\mathcal{V}$). Now, if we compute
$\mu_{\mathcal{V}}(\boldsymbol{\mathcal{C}},\boldsymbol{\mathcal{L}})$
for all $\boldsymbol{\mathcal{C}}\in\mathcal{V}$, then
the application of corollary~\ref{cor:cor2}
implies that if $\boldsymbol{\mathcal{C}}$ is in $\mathcal{V}$
but not in $\mathcal{W}$ then
$\mu_{\mathcal{V}}(\boldsymbol{\mathcal{C}},\boldsymbol{\mathcal{L}})=0$,
while $\mu_{\mathcal{V}}(\boldsymbol{\mathcal{C}},\boldsymbol{\mathcal{L}})=
\mu_\mathcal{W}(\boldsymbol{\mathcal{C}},\boldsymbol{\mathcal{L}})$
otherwise. In other words, if a cluster $\boldsymbol{\mathcal{C}}$
is not the intersection of maximal clusters, then it does not
contribute to the cluster expansion of the free-energy functional.
So, we can constrain this cluster expansion to $\mathcal{W}$ without
loss of generality, as we have indeed done.

\References

\item[] Aigner M 1979 \textit{Combinatorial Theory} (New York: 
	Springer-Verlag) 
\item[] Bowman D R and Levin K 1982 \PR B \textbf{25} 3438--41
\item[] Choudhury N, Patra C N and Ghosh 2002 \JPCM {\bf 14} 11955--63
\item[] Cuesta J A and Mart\'{\i}nez-Rat\'on Y 1997a \PRL {\bf 78} 3681--4
\item[] \dash 1997b \JCP {\bf 107} 6379--89
\item[] Cuesta J A, Mart\'{\i}nez-Rat\'on Y and Tarazona P 2002
        \JPCM \textbf{14} 11965--80
\item[] Denton A R and Ashcroft N W 1991 \PR A {\bf 44} 8242--8
\item[] Evans R 1992 \textit{Fundamentals of Inhomogeneous Fluids} 
	ed~D~Henderson (Dordrecht: Kluwer) pp~85--175
\item[] Kikuchi R 1951 \PR {\bf 81} 988--1003
\item[] Lafuente L 2004 Ph.~D.~thesis \textit{Funcionales de la
	Densidad para Modelos de Red} (Madrid: Universidad Carlos III)
\item[] Lafuente L and Cuesta J A 2002 \JPCM \textbf{14} 12079--97
\item[] \dash 2003 \PR E \textbf{68} 066120
\item[] \dash 2004 \PRL \textbf{93} 130603
\item[] Morita T 1994 \textit{Prog.\ Theor.\ Phys.\ Suppl.} 
	\textbf{115} 27--39
\item[] Rosenfeld Y 1989 \PRL {\bf 63} 980--3
\item[] Rosenfeld Y, Schmidt M, L\"owen H and Tarazona P 1996 \JPCM
        {\bf 8} L577--81
\item[] \dash 1997 \PR E {\bf 55} 4245--63
\item[] Rota G C 1964 \textit{Z.~Wahrscheinlichkeitsrechnung u.~verw.~Geb.}
	\textbf{2} 340--68
\item[] Stanley R P 1999 \textit{Enumerative Combinatorics} vol~1
	(Cambridge: Cambridge University Press)
\item[] Tarazona P 2000 \PRL \textbf{84} 694--7
\item[] Tarazona P 2002 \emph{Physica} A \textbf{306} 243--50
\item[] Tarazona P and Rosenfeld Y 1997 \PR E \textbf{55} R4873--6
\endrefs

\end{document}